\documentclass[pdftex,twocolumn,epjc3]{svjour3}          % twocolumn
\RequirePackage[T1]{fontenc}
\usepackage[utf8x]{inputenc}
\smartqed  % flush right qed marks, e.g. at end of proof
\usepackage{datetime}
\usepackage{mathtools, cuted}
\usepackage{dblfloatfix,float}
\usepackage{amsmath,epsfig,amssymb,multirow}
\RequirePackage{graphicx}
\RequirePackage{mathptmx}      % use Times fonts if available on your TeX system
\RequirePackage{flushend}
\usepackage{cite}
\RequirePackage[numbers,sort&compress]{natbib}
\RequirePackage[colorlinks,citecolor=blue,urlcolor=blue,linkcolor=blue]{hyperref}
\let\oldhref\href
\renewcommand{\href}[2]{\oldhref{#1}{\hbox{#2}}}

\journalname{Eur. Phys. J. C}
\allowdisplaybreaks

\def\wtil#1{\widetilde{#1}}
\begin{document}
\title{Constraining anomalous gauge boson couplings in 
	{\boldmath $e^+e^-\to W^+W^-$} using polarization asymmetries with polarized beams }
%%%%%%%%%%%%%%%%%%%%%%%%%%%%%%%%%%%%%%%%%%%%%%%%%%%%%%%%%%%%%%%%%%%%%%%%%%%%%%
\author{Rafiqul Rahaman\thanksref{e1,adrss} \and 
Ritesh K. Singh\thanksref{e2,adrss}}
\thankstext{e1}{email:rr13rs033@iiserkol.ac.in}
\thankstext{e2}{email:ritesh.singh@iiserkol.ac.in}
\institute{Department of Physical Sciences,
	Indian Institute of Science Education and Research Kolkata,
	Mohanpur, 741246, India\label{adrss}}

\date{Received: date /Accepted: date}

%%%%%%%%%%%%%%%%%%%%%%%%%%%%%%%%%%%%%%%%%%%%%%%%%%%%%%%%%%%%%%%%%%%%%%%%%%%%%%%
\maketitle

\abstract{
	We study the anomalous $W^+W^-V$ ($V=\gamma,Z$) couplings 
	in $e^+e^-\to W^+W^-$  using the complete set of polarization observables of
	 $W$ boson with longitudinally polarized beams.
	We use most general Lorentz invariant form factors parametrization as well as 
	$SU(2)\times U(1)$  invariant dimension $6$ effective operators for the
	effective $W^+W^-V$ couplings. We estimate simultaneous limits on the anomalous
	couplings in both the parametrizations 
	using cross section, forward backward asymmetry and polarization observables of $W$ boson
	with different kinematical cuts using Markov-Chain--Monte-Carlo (MCMC) method 
	for an $e^+e^-$ collider running at centre of mass energy of $\sqrt{s}=500$ GeV
	and ${\cal L}=100$ fb$^{-1}$. The best limits on form factors are obtained 
	to be $1 \sim 5 \times 10^{-2}$ for $e^-$ and $e^+$ polarization being 
	$(+0.4,-0.4)$.
	For operator's coefficients, the best limits are obtained to be $1\sim 16$
	TeV$^{-2}$.
} 
%%%%%%%%%%%%%%%%%%%%%%%%%%%%%%%%%%%%%%%%%%%%%%%%%%%%%%%%%%%%%%%%%%%%%%%%%%%%%%%%%%%%%%%%%%%%
%%%%%%%%%%%%%%%%%%%%%%%%%%%%%%%%%%%%%%%%%%%%%%%%%%%%%%%%%%%%%%%%%%%%%%%%%%%%%%%%%%%%%%%%%%%%
\section{Introduction}

The non-abelian gauge symmetry $SU(2)\times U(1)$ of the Standard Model (SM)
allows the $WWV$ ($V=\gamma,Z$) couplings after the Electro-Weak  
Symmetry Breaking (EWSB) by Higgs field, discovered recently at 
LHC~\cite{Chatrchyan:2012xdj}. To test the EWSB, the $WWV$ couplings have to
be measured precisely, which is still lacking. We intend to study the measurement of
 the couplings using  polarization observables of spin-$1$ 
 boson~\cite{Bourrely:1980mr,Abbiendi:2000ei,Ots:2004hk,Boudjema:2009fz,
 Aguilar-Saavedra:2015yza,Rahaman:2016pqj,Nakamura:2017ihk}.
To test the SM $WWV$ couplings
one has to consider beyond the SM (BSM) couplings in the theory and make sure they do not
appear at all or severely constrained. One way is to  consider $SU(2)\times U(1)$
invariant  higher dimension effective
operators which provide  the $WWV$ form factors after EWSB~\cite{Buchmuller:1985jz} and
add to the SM Lagrangian as
\begin{equation}
{\cal L}_{eft} = {\cal L}_{SM} + \sum_i \frac{c_i^{\cal O}}{\Lambda^2}{\cal O}_i.
\label{eq:eft}
\end{equation}
Here $c_i^{\cal O}$ are  couplings of the dimension-six operators ${\cal O}_i$
and $\Lambda$ is the energy
scale below which the theory is valid.
To the lowest order 
(upto dimension-$6$) the operators contributing to  $WWV$ couplings are~\cite{Hagiwara:1993ck,Degrande:2012wf} 
\begin{eqnarray}\label{eq:opertaors1}
{\cal O}_{WWW}&=&\mbox{Tr}[W_{\mu\nu}W^{\nu\rho}W_{\rho}^{\mu}],\nonumber\\
{\cal O}_W&=&(D_\mu\Phi)^\dagger W^{\mu\nu}(D_\nu\Phi)\label{eq:OW},\nonumber\\
{\cal O}_B&=&(D_\mu\Phi)^\dagger B^{\mu\nu}(D_\nu\Phi),\nonumber\\
{\cal O}_{\wtil{WWW}}&=&\mbox{Tr}[{\tilde W}_{\mu\nu}W^{\nu\rho}W_{\rho}^{\mu}],\nonumber\\
{\cal O}_{\wtil W}&=&(D_\mu\Phi)^\dagger {\tilde W}^{\mu\nu}(D_\nu\Phi),
\label{eq:opertaors5}
\end{eqnarray}
where $\Phi$ is the Higgs doublet field and
\begin{eqnarray}
D_\mu &=& \partial_\mu + \frac{i}{2} g \tau^I W^I_\mu + \frac{i}{2} g^\prime B_\mu,\nonumber \\
W_{\mu\nu} &=& \frac{i}{2} g\tau^I (\partial_\mu W^I_\nu - \partial_\nu W^I_\mu
+ g \epsilon_{IJK} W^J_\mu W^K_\nu ),\nonumber\\
B_{\mu \nu} &=& \frac{i}{2} g^\prime (\partial_\mu B_\nu - \partial_\nu B_\mu).
\end{eqnarray}
Here $g$ and $g^\prime$ are $SU(2)$ and $U(1)$ couplings, respectively.
 Among these operators ${\cal O}_{WWW}$,  ${\cal O}_W$ and  ${\cal O}_B$ are
$CP$ conserving, while ${\cal O}_{\wtil{WWW}}$  and ${\cal O}_{\wtil W}$ are
$CP$ violating. 
 These effective operators in Eq.~\ref{eq:opertaors5},
after EWSB, also provides $ZZV$, $HZV$ couplings which can be examine in various 
processes, e.g. $ZV$ production, $WZ$ production, $HV$ production processes. The couplings
in these processes may contains some other effective operator  as well.

The other way to go beyond the  SM $WWV$ structure is to consider   a
most  general Lorentz 
invariant structure in a model independent way. 
 A Lagrangian corresponding
to the most general Lorentz invariant set of form factors for the $WWV$ couplings is
given by~\cite{Hagiwara:1986vm}
\begin{eqnarray}
{\cal L}_{WWV} &=&ig_{WWV}\left(g_1^V(W_{\mu\nu}^+W^{-\mu}-
W^{+\mu}W_{\mu\nu}^-)V^\nu\right.\nonumber\\
&&\left.
+ig_4^VW_\mu^+W^-_\nu(\partial^\mu V^\nu+\partial^\nu V^\mu)\right.\nonumber\\
&&\left.-ig_5^V\epsilon^{\mu\nu\rho\sigma}(W_\mu^+\partial_\rho W^-_
\nu-\partial_\rho W_\mu^+W^-_\nu)V_\sigma
\right.\nonumber\\
&&\left.
+\frac{\lambda^V}{M_W^2}W_\mu^{+\nu}W_\nu^{-\rho}V_\rho^{\mu}
+\frac{\wtil{\lambda^V}}{M_W^2}W_\mu^{+\nu}W_\nu^{-\rho}\wtil{V}_\rho^{\mu}
\right.\nonumber\\
&&\left.
+\kappa^V W_\mu^+W_\nu^-V^{\mu\nu}+\wtil{\kappa^V}W_\mu^+W_\nu^-\wtil{V}^{\mu\nu}
\right),
\label{eq:Lagrangian}
\end{eqnarray}
where $W_{\mu\nu}^\pm = \partial_\mu W_\nu^\pm - 
\partial_\nu W_\mu^\pm$, $V_{\mu\nu} = \partial_\mu V_\nu - 
\partial_\nu V_\mu$, 
$\wtil{V}^{\mu\nu}=1/2\epsilon^{\mu\nu\rho\sigma}V_{\rho\sigma}$,
and the overall coupling constants are defined as
$g_{WW\gamma}=-g\sin\theta_W$ and $g_{WWZ}=-g\cos\theta_W$, $\theta_W$ being the weak
mixing angle. In the SM
$g_1^V=1$, $\kappa^V=1$ and other couplings are zero. The anomalous part in $g_1^V$,
 $\kappa^V$ would be $\Delta g_1^V=g_1^V-1$, $\Delta\kappa^V=\kappa^V-1$, respectively. 
 The couplings $g_1^V$, $\kappa^V$
and $\lambda^V$ of Eq.~\ref{eq:Lagrangian} conserve $CP$ (both $C$ and $P$-even), 
while $g_4^V$ (odd in $C$, even in $P$), $\wtil{\kappa^V}$
and $\wtil{\lambda^V}$  (even in $C$, odd in $P$) violate $CP$. On the other hand 
$g_5^V$ violates both $C$ and $P$ leaving it to $CP$ conserving. We label these set of  $14$
anomalous couplings to be $c_i^{\cal L}$ as given in Eq.~\ref{eq:ciL} in~\ref{apendix:a} for 
later uses.

On restricting to the  $SU(2)\times U(1)$ gauge, the coupling ($c_i^{\cal L}$) of the Lagrangian
in Eq.~\ref{eq:Lagrangian} can be written in terms of  the couplings of the operators
 in Eq.~\ref{eq:opertaors5} as~\cite{Hagiwara:1993ck,Wudka:1994ny,Degrande:2012wf}

\begin{eqnarray}
\Delta g_1^Z & = & c_W\frac{M_Z^2}{2\Lambda^2},\nonumber\\
g_4^V &=& g_5^V=g_1^\gamma=0,\nonumber\\
\lambda_\gamma & = & \lambda_Z = c_{WWW}\frac{3g^2M_W^2}{2\Lambda^2},\nonumber\\
\wtil{\lambda}_\gamma & = & \wtil{\lambda}_Z = c_{\wtil{WWW}}\frac{3g^2M_W^2}{2\Lambda^2},\nonumber\\
\Delta\kappa_\gamma & = & (c_W+c_B)\frac{M_W^2}{2\Lambda^2},\nonumber\\
\Delta\kappa_Z & = & (c_W-c_B\tan^2\theta_W)\frac{M_W^2}{2\Lambda^2},\nonumber\\
\wtil{\kappa}_\gamma & = &
c_{\wtil{W}}\frac{M_W^2}{2\Lambda^2},\nonumber\\
\wtil{\kappa}_Z & = &
-c_{\wtil{W}}\tan^2\theta_W\frac{M_W^2}{2\Lambda^2}.
\label{eq:Operator-to-Lagrangian}
\end{eqnarray}
In this case some of the Lagrangian couplings become dependent to each others and 
they are 
\begin{eqnarray}
&&\Delta g_1^Z=\Delta \kappa_Z + \tan^2\theta_W \Delta \kappa_\gamma,\nonumber\\
&&\wtil \kappa_Z + \tan^2\theta_W \tilde \kappa_\gamma=0.
\end{eqnarray}
We label the non-vanishing $9$ couplings in $SU(2)\times U(1)$ gauge as $c_i^{{\cal L}_g}$
 given in Eq.~\ref{eq:ciLg} in \ref{apendix:a} for later uses.

The anomalous $WWV$ couplings has been studied in the effective operators  formalism as well
as in the effective vertex factor approach given 
in the Lagrangian ${\cal L}_{WWV}$ (Eq.~\ref{eq:Lagrangian}) in $SU(2)\times U(1)$ gauge for $e^+$-$e^-$ linear 
collider~\cite{Gaemers:1978hg,Hagiwara:1986vm,Bilchak:1984ur,Hagiwara:1992eh,
Wells:2015eba,Buchalla:2013wpa,Zhang:2016zsp,Berthier:2016tkq,Bian:2015zha,Bian:2016umx,
Choudhury:1996ni,Choudhury:1999fz}, Large Hadron electron collider (LHeC)
~\cite{Biswal:2014oaa,Cakir:2014swa}, $e$-$\gamma$ collider~\cite{Kumar:2015lna}, 
hadron collider 
(LHC)~\cite{Baur:1987mt,Dixon:1999di,Falkowski:2016cxu,Azatov:2017kzw,Bian:2015zha,
Bian:2016umx,Butter:2016cvz,Baglio:2017bfe,Li:2017esm}. These couplings has also
been addressed from loop level contribution~\cite{Hagiwara:1992eh} and  Georgi-Machacek 
model~\cite{Arroyo-Urena:2016gjt}. Some $CP$-violating $WWV$ couplings has been studied
in Ref.s~\cite{Choudhury:1999fz,Li:2017esm}. 

On the experimental side the anomalous $WWV$ couplings have been explored and stringent limits
on them  have been obtained 
in  different process ( $W^\pm V$, $W^\pm jj$ production) and 
different channel ($l\nu_lJ,~qql\nu_l $) in the LEP~\cite{Abbiendi:2000ei,Abbiendi:2003mk,
Abdallah:2008sf,Schael:2013ita}, Tevatron~\cite{Aaltonen:2007sd,Abazov:2012ze},  LHC~\cite{Aaboud:2017cgf,
Sirunyan:2017bey,Aaboud:2017fye,Khachatryan:2016poo,
Aad:2016ett,Aad:2016wpd,Chatrchyan:2013yaa,
RebelloTeles:2013kdy,ATLAS:2012mec,Chatrchyan:2012bd,Aad:2013izg,
Chatrchyan:2013fya}, Tevatron-LHC~\cite{Corbett:2013pja}. The  tightest one parameter limit
observed on the  anomalous couplings from   experiments are given in 
Table~\ref{tab:aTGC_constrain_form_collider}. 
The tightest limit on operator couplings ($c_i^{\cal O}$) are obtained in Ref.~ 
\cite{Sirunyan:2017bey} for $CP$-even
ones and in Ref.~\cite{Aaboud:2017fye} for $CP$-odd ones. The limits on the couplings of the 
Lagrangian in Eq.~\ref{eq:Lagrangian} are tighter when $SU(2)\times U(1)$ symmetry is assumed  
and the tightest ones are obtained in Ref.~\cite{Sirunyan:2017bey} for $CP$-even
 and in Ref.~\cite{Aaboud:2017fye,Abdallah:2008sf} for $CP$-odd parameters. These limits on 
$c_i^{{\cal L}_g}$  are actually translated from the limits of the operator couplings $c_i^{\cal O}$. The tightest limits on the  couplings, which are zero when $SU(2)\times U(1)$ symmetry is assumed (see Eq.~\ref{eq:Operator-to-Lagrangian}),
are obtained  in Ref.~\cite{Abdallah:2008sf,Abbiendi:2003mk} considering the Lagrangian in 
Eq.~\ref{eq:Lagrangian}. 

\begin{table*}[!ht]
	\centering
	\caption{\label{tab:aTGC_constrain_form_collider} The list of tightest limits observed on 
		anomalous couplings of Eq~(\ref{eq:opertaors5}) (dimension-$6$ operators),  Eq.~(\ref{eq:Lagrangian}) (effective Lagrangian)
		in $S(2)\times U(1)$ (except $g_4^Z$ and $g_5^Z$) at $95\%$ C.L.  from experiments}
	\renewcommand{\arraystretch}{1.5}
	\begin{tabular*}{\textwidth}{@{\extracolsep{\fill}}lll@{}}\hline
				$c_i^{\cal O}$	        & Limits (TeV$^{-2}$)   & Remark\\\hline 
				$\frac{c_{WWW}}{\Lambda^2}$	                & $[-2.7,+2.7]$ &CMS $\sqrt{s}=8$ TeV, ${\cal L}=19$ fb$^{-1}$, $SU(2)\times U(1)$~\cite{Sirunyan:2017bey} \\
				$\frac{c_{W}}{\Lambda^2}$                     & $[-2.0,+5.7]$ &CMS~\cite{Sirunyan:2017bey} \\
				$\frac{c_{B}}{\Lambda^2}$	                & $[-14,+17]$ &CMS~\cite{Sirunyan:2017bey} \\    
				$ \frac{c_{\widetilde{WWW}}}{\Lambda^2}$    &$[-11,+11]$  &ATLAS $\sqrt{s}=7(8)$ TeV, ${\cal L}=4.7(20.2)$ fb$^{-1}$ ~\cite{Aaboud:2017fye}\\
				$ \frac{c_{\widetilde{W}}}{\Lambda^2}$      &$[-580,580]$  &ATLAS~\cite{Aaboud:2017fye} \\
				\hline
		$c_i^{{\cal L}_g}$ & Limits ($\times 10^{-2}$) & Remark\\ \hline
		$\lambda^V$ &  $[-1.1,+1.1]$ &CMS~\cite{Sirunyan:2017bey}\\
		$\Delta\kappa^\gamma$ &$[-4.4,+6.3]$&CMS~\cite{Sirunyan:2017bey}\\
		$\Delta g_1^Z$ & $[-0.87,+2.4]$ &  CMS~\cite{Sirunyan:2017bey}\\
		$\Delta\kappa^Z$ & $[-0.5,+0.4]$ &CMS~\cite{Sirunyan:2017bey}\\
		$\wtil{\lambda^V}$ & $[-4.7,+4.6]$ &ATLAS~\cite{Aaboud:2017fye}\\
		$\wtil{\kappa^Z}$  & $[-14,-1]$ & DELPHI (LEP2), $\sqrt{s}=189$-$209$ GeV, ${\cal L}=520$ pb$^{-1}$~\cite{Abdallah:2008sf}\\
		\hline
		$c_i^{{\cal L}}$  & Limits ($\times 10^{-2}$) & Remark\\ \hline
				$g_4^Z$ & $[-59,-20]$ &DELPHI~\cite{Abdallah:2008sf}\\ 
				$g_5^Z$  & $[-16,+9.0]$ &OPAL (LEP), $\sqrt{s}=183$-$209$ GeV, ${\cal L}=680$ pb$^{-1}$~\cite{Abbiendi:2003mk} \\
				\hline
	\end{tabular*}
\end{table*}

The process $e^+e^-\to W^+W^-$ will be one of the important process which will be studied
at the future International Linear Collider (ILC)~\cite{Djouadi:2007ik,Baer:2013cma,
Behnke:2013xla} for precision test~\cite{MoortgatPick:2005cw} as well as for BSM physics. This process has been studied earlier for SM phenomenology
as well as for various BSM physics with and without beam polarizations ~\cite{Hagiwara:1986vm,Gounaris:1992kp,Ananthanarayan:2009dw,Ananthanarayan:2010bt,Ananthanarayan:2011ga,Andreev:2012cj}.
 Here we intend to study $WWV$ anomalous couplings in $e^+e^-\to W^+W^-$ at $\sqrt{s}=500$ GeV,
 ${\cal L}=100$ fb$^{-1}$ using the cross section, forward backward asymmetry and $8$ 
polarizations asymmetries  of $W^-$ with longitudinally  polarized $e^+$ and $e^-$ beams. 
Here, first we study the anomalous couplings ($c_i^{\cal L}$) in the 
Lagrangian ${\cal L}_{WWV}$ ( Eq.~\ref{eq:Lagrangian})
 and estimate simultaneous limits on all $14$ couplings. Next we consider the $SU(2)\times U(1)$
 effective operators (given in Eq.~\ref{eq:opertaors5}) contributing to the anomalous couplings and obtain
 simultaneous limit on the corresponding couplings ($c_i^{\cal O}$) given in Eq.~\ref{eq:ciO}. 
 The translated limit on the reaming couplings ($c_i^{{\cal L}_g}$) given in Eq.~\ref{eq:ciLg} 
 of the Lagrangian in Eq.~\ref{eq:Lagrangian} has also been obtained.

The rest of the paper is arranged in the following way. In Sect.~\ref{sec:2} we introduce 
the compete set polarization observables of a spin-$1$ particle along with 
the forward backward asymmetry and study the effect of beam polarizations on the observables.
In Sect.~\ref{sec:3} we use the vertex form factors for the Lagrangian in Eq.~\ref{eq:Lagrangian}
and obtained expressions for all the observables. In this section
we cross check analytical results against the numerical result from 
{\tt MadGraph5}~\cite{Alwall:2014hca} for sanity checking. We also study the  $\cos\theta$ (of $W$ )
dependences  of the observables  and study their sensitivity on the anomalous couplings. 
In this section we also estimates simultaneous limits on $c_i^{{\cal L}}$,   
 $c_i^{\cal O}$  and the translated limits on
$c_i^{{\cal L}_g}$. Next in  Sect.~\ref{sec:4} we give insight 
 on the choice of beam polarizations in this process. We conclude in Sect.~\ref{sec:conclusion}.

%%%%%%%%%%%%%%%%%%%%%%%%%%%%%%%%%%%%%%%%%%%%%%%%%%%%%%%%%%%%%%%%%%%%%%%%%%%%%%%%%%%%%%%%%%%%%%%%%%%%
%%%%%%%%%%%%%%%%%%%%%%%%%%%%%%%%%%%%%%%%%%%%%%%%%%%%%%%%%%%%%%%%%%%%%%%%%%%%%%%%%%%%%%%%%%%%%%%%%%%
\begin{figure}[!b]
\centering
\includegraphics[width=8.50cm]{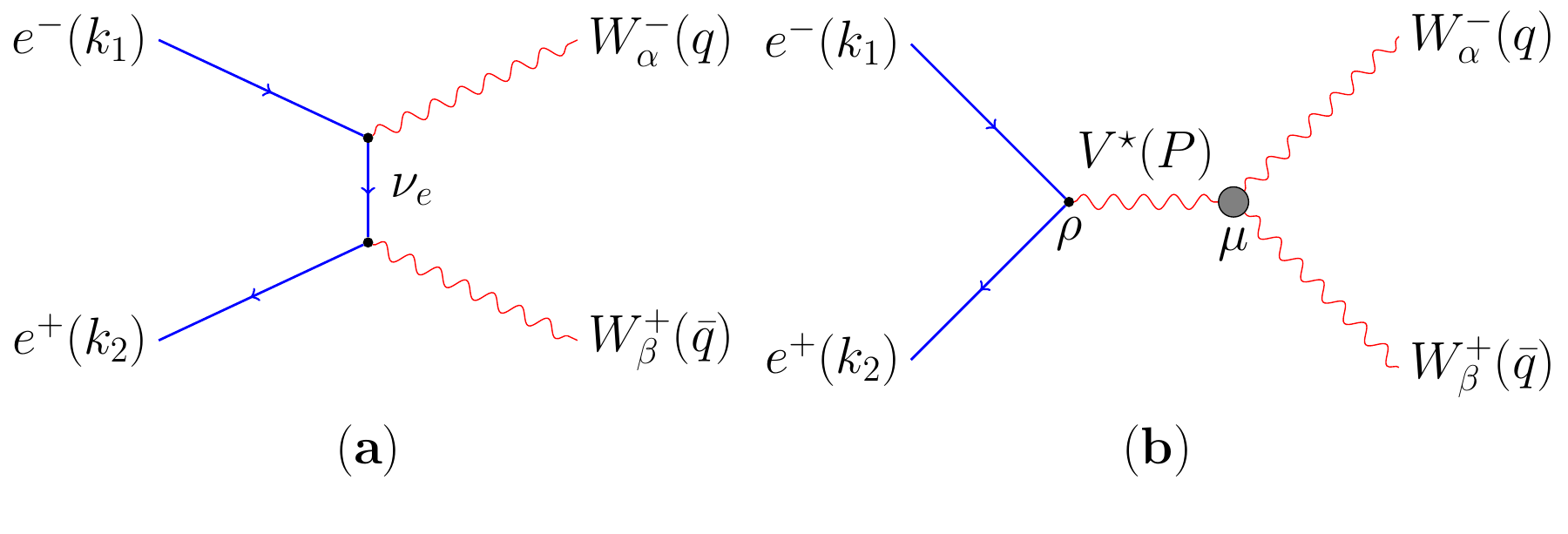}
\caption{\label{fig:Feynman-ee-ww}Feynman diagram of $e^+e^-\to W^+W^-$,
	(a) $t$-channel and (b) $s$-channel with anomalous $W^+W^-V$ 
	($V=\gamma,Z$) vertex contribution shown as blob} 
\end{figure}

\begin{figure*}[!ht]
	\centering
	\includegraphics[height=6.70cm]{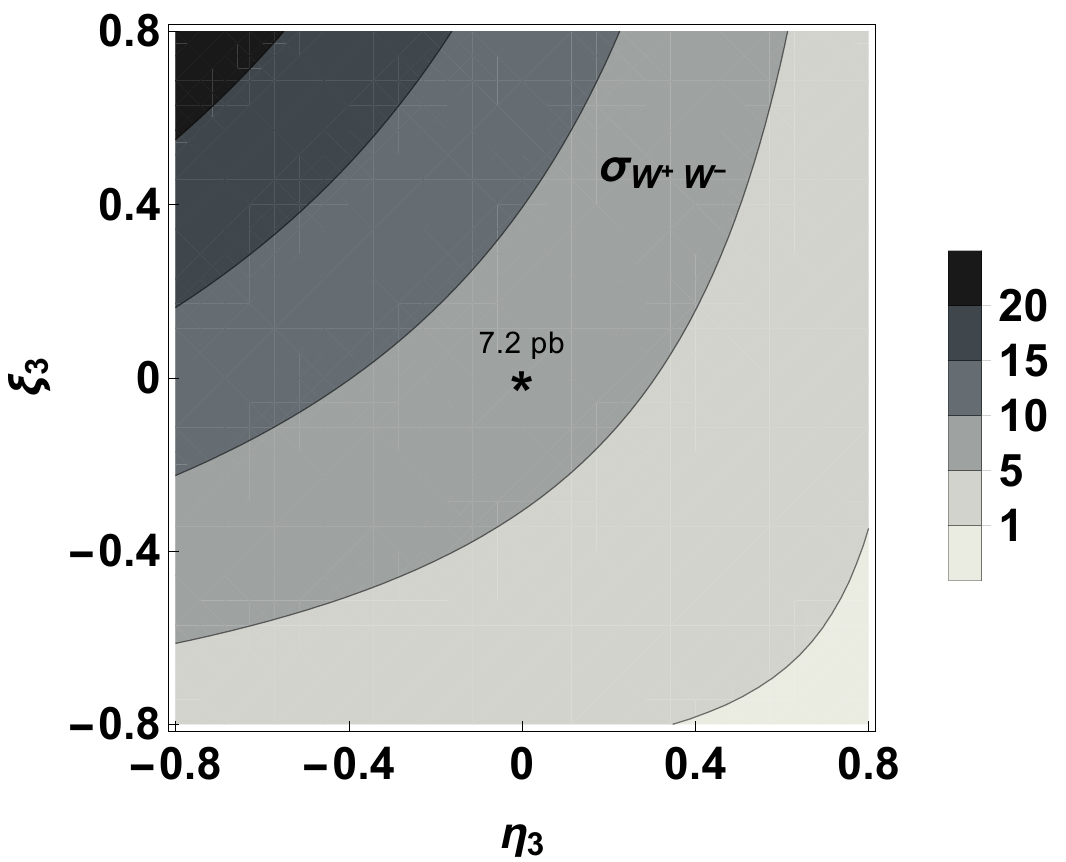}
	\includegraphics[height=6.70cm]{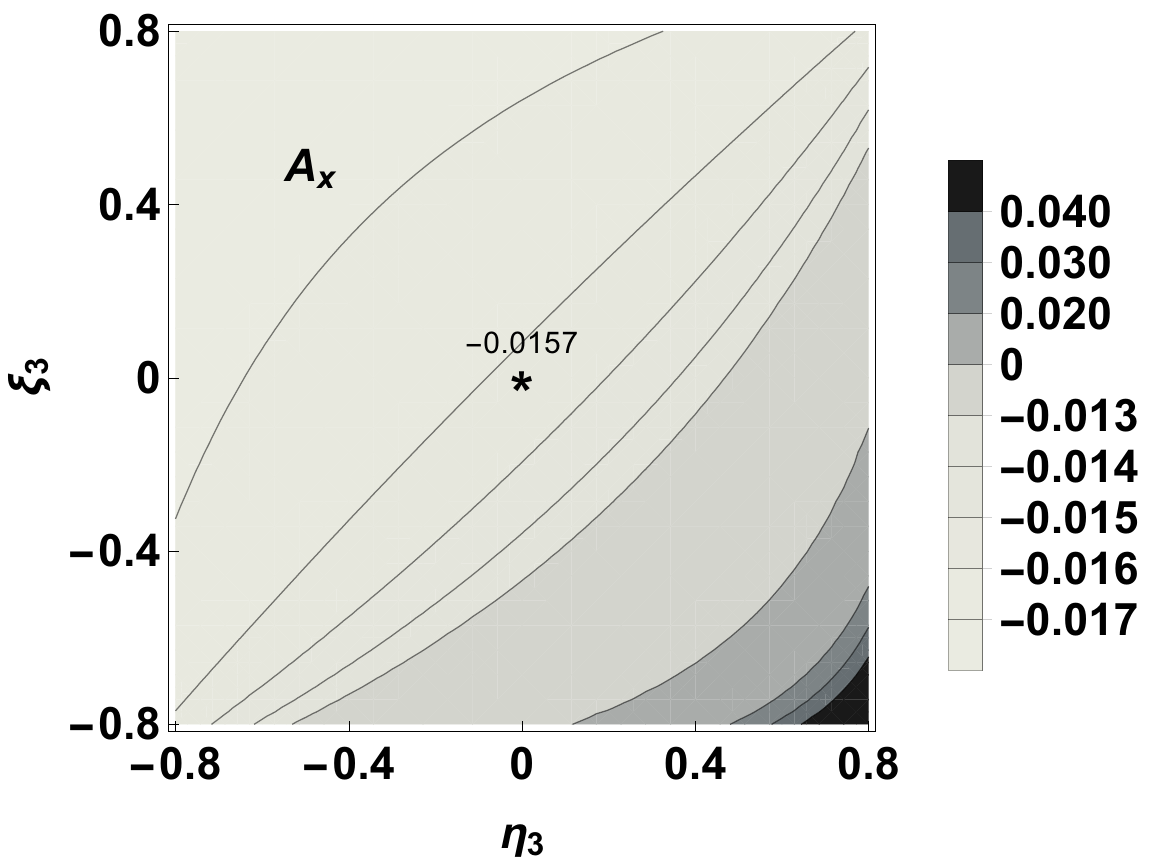}
	\caption{\label{fig:Sigma_and_Afb_eta3xi3}  Production cross section $\sigma_{W^+W^-}$ in pb (left)
		and polarization asymmetry $A_x$ (right) in the SM as a function of longitudinal beam polarization
		$\eta_3$ (for $e^-$) and $\xi_3$ (for $e^+$)  in $e^+e^-\to W^+W^-$ at $\sqrt{s}=500$ GeV. 
		The asterisk represent the unpolarized point and the number near it corresponds to the SM
		 values for corresponding observables for unpolarized beams} 
\end{figure*}
\section{Observables and effect of beam polarizations on them}\label{sec:2}

We study $W^+W^-$ production at ILC running at $\sqrt{s}=500$ GeV and 
integrated luminosity ${\cal L}=100$ fb$^{-1}$ using longitudinal polarization  
of $e^-$ and $e^+$ beams. The Feynman diagram for the process is  shown in 
Fig.~\ref{fig:Feynman-ee-ww} where Fig.~\ref{fig:Feynman-ee-ww}\textbf{a} corresponds
to the $\nu_e$ mediated  $t$-channel diagram and the 
Fig.~\ref{fig:Feynman-ee-ww}\textbf{b} corresponds to the $V~(Z,\gamma)$ mediated $s$-channel 
  diagram. The decay mode is chosen to be
 \begin{equation}
 W^+\to q_u~\bar{q}_d   ,~~~~~  W^-\to l^-~\bar{\nu}_l, 
 \end{equation}
where $q_u$ and $q_d$ are up-type and down-type quarks, respectively.
We use complete set of eight spin-$1$  observables of $W^-$ boson
~\cite{Aguilar-Saavedra:2015yza,Rahaman:2016pqj}.

The $W$ boson being a spin-$1$ particle, its normalised production density matrix
in the spin basis can be written as~\cite{Bourrely:1980mr,Boudjema:2009fz}
\begin{equation}\label{eq:spin-desnity-matrix}
\rho(\lambda,\lambda^\prime)=\dfrac{1}{3}\Bigg[I_{3\times 3} +\dfrac{3}{2} \vec{p}.\vec{S}
+\sqrt{\dfrac{3}{2}} T_{ij}\big(S_iS_j+S_jS_i\big) \Bigg],
\end{equation}
where $\vec{p}=\{p_x,p_y,p_z\}$ is the vector polarization of a spin-$1$ particle,
 $\vec{S}=\{S_x,S_y,S_z\}$ are the spin basis and $T_{ij}$
is the $2^{nd}$-rank symmetric traceless tensor, $\lambda$ and $\lambda^\prime$ are
helicities of the particle. The tensor $T_{ij}$ has $5$ independent elements, which are
$T_{xy}$, $T_{xz}$, $T_{yz}$, $T_{xx}-T_{yy}$ and $T_{zz }$. Combining the $\rho(\lambda,\lambda^\prime)$ with   normalised
decay density matrix of the  particle  to a pair of fermion $f$, 
the differential cross section  would be~\cite{Boudjema:2009fz}
 \begin{eqnarray}
 \frac{1}{\sigma} \ \frac{d\sigma}{d\Omega_f} &=&\frac{3}{8\pi} \left[
 \left(\frac{2}{3}-(1-3\delta) \ \frac{T_{zz}}{\sqrt{6}}\right) + \alpha \ p_z
 \cos\theta_f \right.\nonumber\\
 &+& \sqrt{\frac{3}{2}}(1-3\delta) \ T_{zz} \cos^2\theta_f
 \nonumber\\
 &+&\left(\alpha \ p_x + 2\sqrt{\frac{2}{3}} (1-3\delta)
 \ T_{xz} \cos\theta_f\right) \sin\theta_f \ \cos\phi_f \nonumber\\
 &+&\left(\alpha \ p_y + 2\sqrt{\frac{2}{3}} (1-3\delta)
 \ T_{yz} \cos\theta_f\right) \sin\theta_f \ \sin\phi_f \nonumber\\
 &+&(1-3\delta) \left(\frac{T_{xx}-T_{yy}}{\sqrt{6}} \right) \sin^2\theta_f
 \cos(2\phi_f)\nonumber\\
 &+&\left. \sqrt{\frac{2}{3}}(1-3\delta) \ T_{xy} \ \sin^2\theta_f \
 \sin(2\phi_f) \right]. \label{eq:angular_distribution}
 \end{eqnarray}
 Here $\theta_f$, $\phi_f$ are the polar and the azimuthal orientation of the  fermion $f$,
 in the rest frame of the particle ($W$) with its would be momentum along  $z$-direction. In this case
 $\alpha=-1$ and $\delta=0$. The vector polarizations $\vec{p}$ and independent tensor
 polarizations $T_{ij}$ are calculable from the asymmetries constructed from the 
 decay angular information of lepton (here $l^-$). For example 
 $p_x$ can be calculated from the asymmetry $A_x$ as 
 \begin{eqnarray}\label{eq:pol_decay_Ax}
 A_x=
\dfrac{\sigma(\cos\phi>0)-\sigma(\cos\phi<0)}{\sigma(\cos\phi>0)+\sigma(\cos\phi<0)}
 \equiv   \frac{3 \alpha  p_x}{4}.
 \end{eqnarray}
  The asymmetries corresponding to all other polarizations,  vector polarizations  
 $p_y,p_z$ and independent tensor polarizations $T_{ij}(i,j=x,y,z)$
 are $A_y$, $A_z$, $A_{xy}$, $A_{xz}$, $A_{yz}$, $A_{x^2-y^2}$, $A_{zz}$ 
(see Ref.~\cite{Rahaman:2016pqj} for details). 

Owing to the 
$t$-channel process (Fig.~\ref{fig:Feynman-ee-ww}\textbf{a}) and not a $u$-channel 
process  like in $ZV$ production~\cite{Rahaman:2016pqj,Rahaman:2017qql},
 the $W^\pm$ produced are not 
forward backward symmetric. We add forward-backward asymmetry defined as 
\begin{equation}
A_{fb}=\frac{1}{\sigma_{W^+W^-}}\Bigg[\int_0^1 \frac{d\sigma_{W^+W^-}}{d\cos\theta} 
-\int_{-1}^0 \frac{d\sigma_{W^+W^-}}{d\cos\theta}    \Bigg]
\end{equation}
of the $W$ to the set of observables making total of ten observables including
the cross section. Here $\cos\theta$ is the production angle of 
the $W^-$ w.r.t. the $e^-$ beam direction and $\sigma_{W^+W^-}$ is the production cross 
section. 

These asymmetries can be measured in a real collider
from the final state lepton $l^-$. One has to calculate the asymmetries 
at the rest frame of $W^-$ which require the missing $\bar{\nu_l}$ momenta to be
reconstructed. At an $e^+$ $e^-$ collider, as studied here, reconstructing the missing 
$\bar{\nu_l}$ is possible because only one missing particle is involved and
 no parton distribution function (PDF)
is involved, i.e., initial momentas are known. But 
for a collider where PDF is involved, reconstructing the actual missing momenta
may not be  possible.

We explore the dependence of the cross section and asymmetries on
longitudinal polarization $\eta_3$ of $e^-$ and $\xi_3$ of $e^+$.   
In Fig.~\ref{fig:Sigma_and_Afb_eta3xi3} we show the production cross section 
$\sigma_{W^+W^-}$  and $A_x$ as a function of beam polarization as an example. 
The cross section decreases along $\eta_3=-\xi_3$ path from $20$ pb on the 
left-top corner to $7.2$ pb at unpolarized point and further to $1$ pb in the 
right-bottom corner. This is because the $W^\pm$ 
couples to left chiral $e^-$ i.e., it requires $e^-$ to be negatively 
polarized and $e^+$ to be positively polarized for higher cross section.
 The variation 
of $A_{fb}$ (not shown) with beam polarization is same as cross section but 
the variation is very slow above the line $\eta_3=\xi_3$. From this we can expect 
that a positive $\eta_3$ and a negative $\xi_3$ will reduce the SM values of 
observables increasing the $S/\sqrt{B}$ ratio ($S=$ signal, $B=$ background).
Some other asymmetries like $A_x$ has opposite dependence on the beam
polarizations compared to the cross section, its modulus reduces for negative $\eta_3$ and positive
$\xi_3$. So, some beam polarization in between  ($\pm 0.8,\mp 0.8$) may come out to be 
a good choice for obtaining best simultaneous limits on anomalous couplings as will
be explored in the next section.

%%%%%%%%%%%%%%%%%%%%%%%%%%%%%%%%%%%%%%%%%%%%%%%%%%%%%%%%%%%%%%%%%%%%%%%%%%%%%%%%%%%%%%%%%%%%%%%%%%%%%%%%%%
%%%%%%%%%%%%%%%%%%%%%%%%%%%%%%%%%%%%%%%%%%%%%%%%%%%%%%%%%%%%%%%%%%%%%%%%%%%%%%%%%%%%%%%%%%%%%%%%%%%%%%%%%%

\begin{figure}
	\centering
	\includegraphics[width=8.50cm]{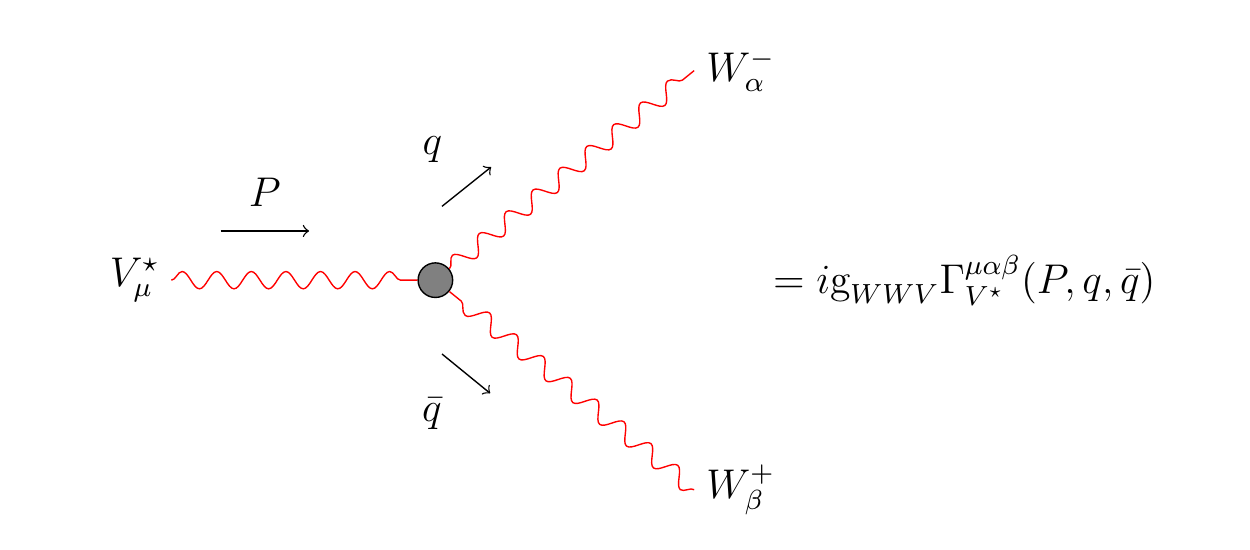}
	\caption{\label{fig:wwv_vertex}The $WWV$ vertex showing anomalous contribution 
represented as blob on top of SM. The momentum $P$ is incoming to the vertex
		while, $q$, $\bar{q}$ are outgoing from the vertex} 
\end{figure}

\section{Probe to the Anomalous Lagrangian}\label{sec:3}
The $W^+W^-V$ vertex (Fig.~\ref{fig:wwv_vertex})
for the Lagrangian in Eq.~\ref{eq:Lagrangian} for on-shell $W$s would be 
$ig_{WWV}\Gamma_V^{\mu\alpha\beta}$ \cite{Gaemers:1978hg,Hagiwara:1986vm}
and it is given by
\begin{eqnarray}
\Gamma_V^{\mu\alpha\beta}&=&f_1^V(q-\bar q)^\mu g^{\alpha\beta}-
\frac{f_2^V}{M_W^2}(q-\bar q)^\mu P^\alpha P^\beta\nonumber\\&&
+f_3^V(P^\alpha g^{\mu\beta}-P^\beta g^{\mu\alpha})\nonumber
+if_4^V(P^\alpha g^{\mu\beta}+P^\beta g^{\mu\alpha})\nonumber\\&&
+if_5^V\epsilon^{\mu\alpha\beta\rho}(q-\bar{q})_\rho
-f_6^V\epsilon^{\mu\alpha\beta\rho}P_\rho \nonumber\\&&
+\frac{\wtil{f_7^V}}{M_W^2}
\left(\bar{q}^\alpha\epsilon^{\mu\beta\rho\sigma} + 
q^\beta\epsilon^{\mu\alpha\rho\sigma}\right)q_\rho\bar{q}_\sigma,
\label{eq:wwv_vertex}
\end{eqnarray}
where $P,q,\bar q$ are the four-momenta of $V,W^-,W^+$, respectively. The 
momentum conventions are shown in  Fig.~\ref{fig:wwv_vertex}. The blob in
the vertex of Figs.~\ref{fig:Feynman-ee-ww} \& \ref{fig:wwv_vertex} %is given to 
represent the presence of anomalous contribution. The form factor $f_i$s has been 
obtained from the Lagrangian in Eq.~\ref{eq:Lagrangian} using  
{\tt FeynRules}~\cite{Alloul:2013bka} to be
\begin{eqnarray}\label{eq:reltn_f_Lagrn}
&&f_1^V=g_1^V + \frac{\hat{s}}{2M_W^2}\lambda^V, \hspace{0.2cm}
f_2^V=\lambda^V,\hspace{0.2cm}
f_3^V=g_1^V + \kappa^V + \lambda^V , \nonumber\\
&&f_4^V=g_4^V,\hspace{0.2cm}
f_5^V=g_5^V,\hspace{0.2cm}
f_6^V=\widetilde{\kappa^V} +
\left(1-\frac{\hat{s}}{2M_W^2} \right)\widetilde{\lambda^V}, \nonumber\\
&&\wtil{f_7^V}=\widetilde{\lambda^V}.
\end{eqnarray}

\begin{figure}
	\centering
	\includegraphics[width=8.50cm]{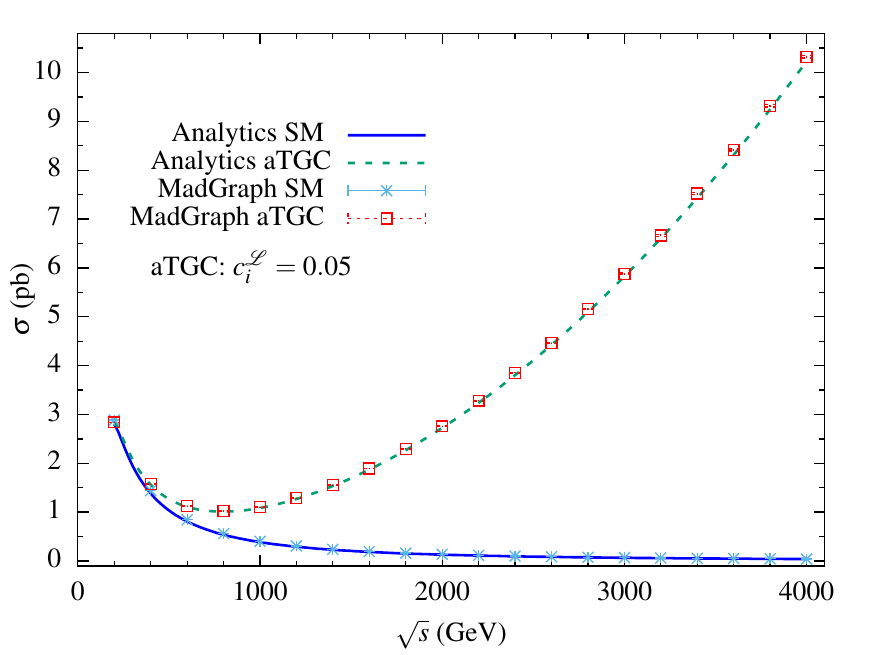}
	\includegraphics[width=8.50cm]{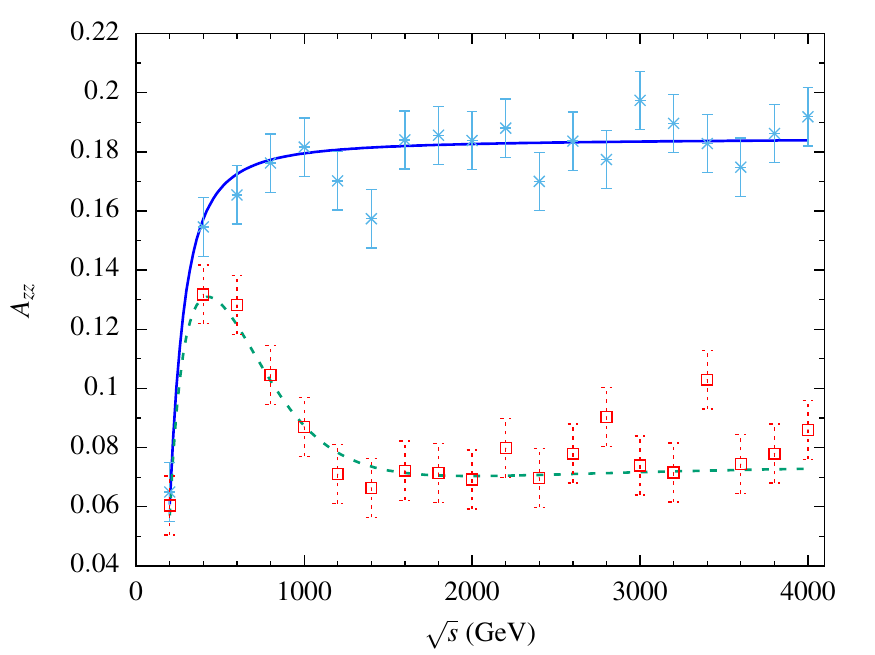}
	\caption{\label{fig:SanityCheck}  The cross section $\sigma$ in pb (above)
and  asymmetry $A_{zz}$ (below) for the SM   and {\tt aTGC} with all anomalous 
couplings  ($c_i^{\cal L}$) at $0.05$ in $e^+e^-\to W^+(\to hadrons)W^-(\to l^-\bar{\nu_l})$  
 as a function of $\sqrt{s}$ for  the SM analytics ({\it solid}/blue) and {\tt aTGC} analytics
 ({\it dashed} /green). The {\it asterisk-solid} (cyan) points and {\it box-dashed} (red) 
points with error-bar correspond to result from {\tt MadGraph5}.
 The errorbar are given for number of events of $10^4$ } 
\end{figure}

We use the vertex factor in Eq.~\ref{eq:wwv_vertex} for the analytical 
calculation of our observables and cross validate them  numerically
with {\tt MadGraph5}~\cite{Alwall:2014hca} implementation
of Eq.~\ref{eq:Lagrangian}. As an example, we present two observables 
$\sigma_{W^+W^-}$ and $A_{zz}$ for the SM ($c_i^{\cal L}=0.0$) and for a 
chosen couplings point $c_i^{\cal L}=0.05$, in Fig.~\ref{fig:SanityCheck}. 
The agreement between the analytical and the numerical calculations over a range
of $\sqrt{s}$ indicates the validity of relations in Eq.~\ref{eq:reltn_f_Lagrn},
specially the $s$ dependence of $f_1^V$ and $f_6^V$.

Analytical  expressions of all the observables has been obtained and their
dependence on the anomalous couplings $c_i^{\cal L}$ are given  in
 Table~\ref{tab:param_dependence} in~\ref{apendix:a}.
The $CP$-even couplings in $CP$-even observables $\sigma$, $A_x$, $A_z$, $A_{xz}$
, $A_{x^2-y^2}$, $A_{zz}$ appear in linear as well as in quadratic form but do not
 appear  in the $CP$-odd observables $A_y$, $A_{xy}$, $A_{yz}$. On the other hand
 $CP$-odd couplings appears linearly in  $CP$-odd observables and quadratically
 in $CP$-even observables. Thus the $CP$-even couplings may have
 double patch in their confidence interval leading to asymmetric limits  
 which will be discussed in Sect.~\ref{sec:3.1}. On the other hand the $CP$-odd
 couplings will have single patch in their confidence interval and will poses 
 symmetric limits. To this end we discuss sensitivity and limits on the anomalous
 couplings in the next subsection.
%-----------------------------------------------------------------------------
\begin{figure*}[htb!]
	\centering
	\includegraphics[width=8.650cm]{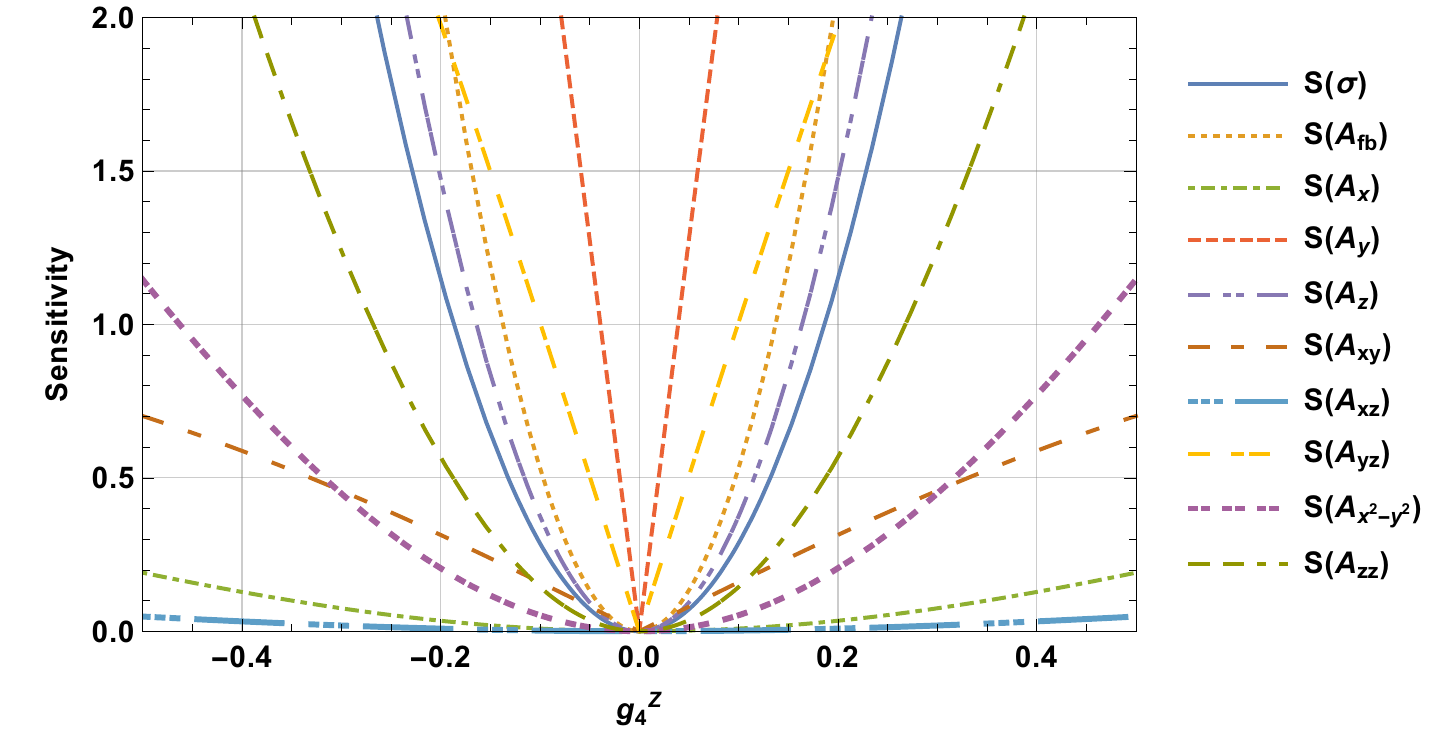}
	\includegraphics[width=8.650cm]{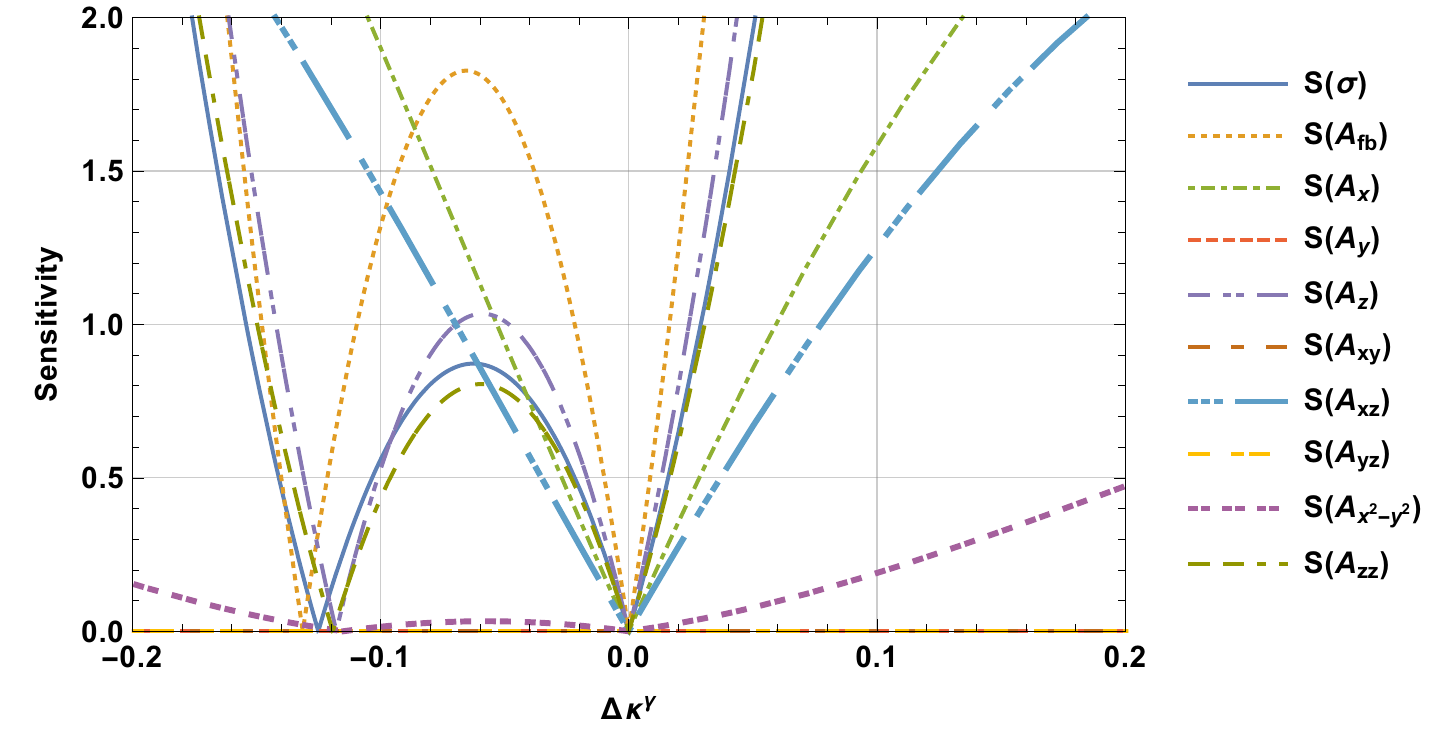}
	\caption{\label{fig:sensitivity} The one parameter sensitivity of  cross section $\sigma$, 
$A_{fb}$ and $8$ polarization asymmetries ($A_i$) on $g_4^Z$ (left)
and on $\Delta\kappa^\gamma$ (right)  in $e^+e^-\to W^+W^-$ at $\sqrt{s}=500$ GeV, ${\cal L}=100$ fb$^{-1}$} 
\end{figure*}
%%%%%%%%%%%%%%%%%%%%%%%%%%%%%%%%%%%%%%%%%%%%%%%%%%%%%%%%%%%%%%%%%%%%%%%%%%%%%%%%%%%%%%%%%%%%%%%%
\subsection{Sensitivity of observables on anomalous couplings and their binning}
\label{sec:3.1}
%----------
\begin{figure*}[!t]
	\centering
	\includegraphics[width=8.50cm]{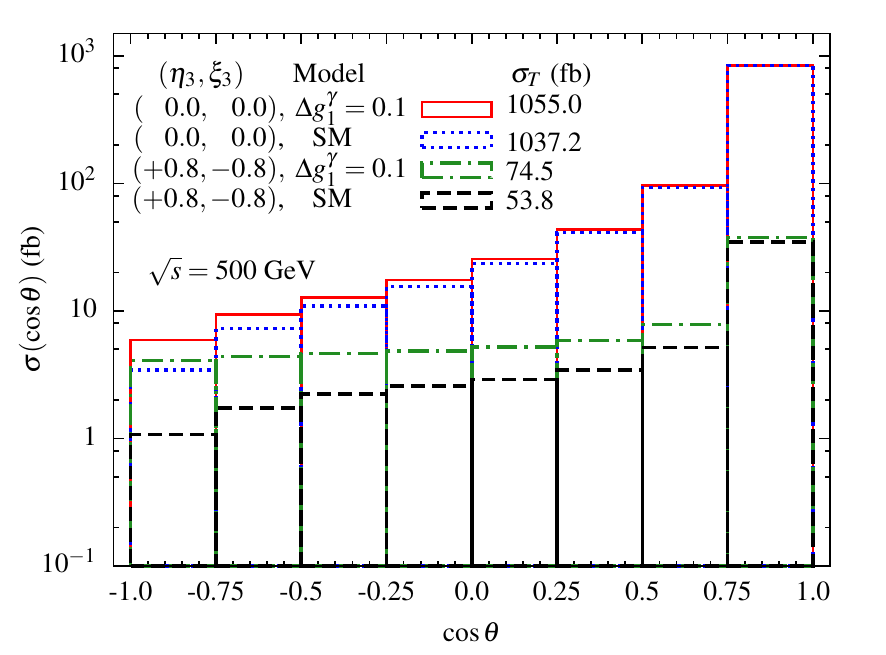}
	\includegraphics[width=8.50cm]{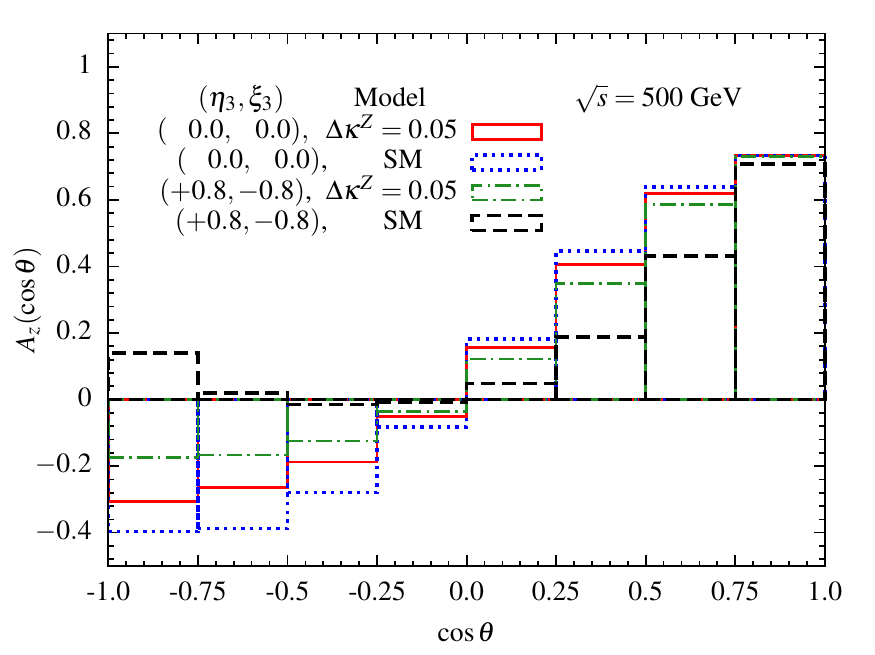}
	\includegraphics[width=8.50cm]{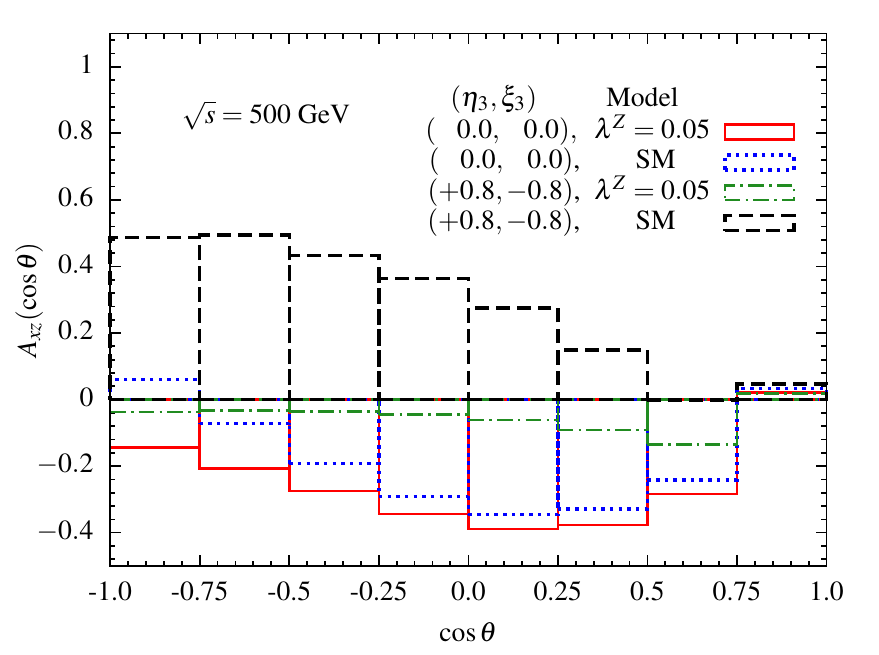}
	\includegraphics[width=8.50cm]{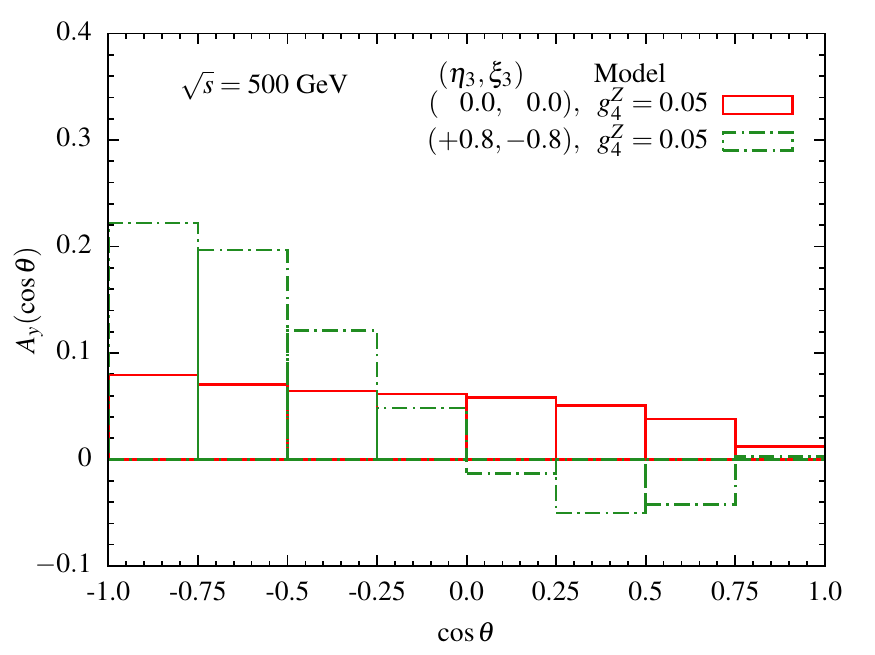}
	\caption{\label{fig:diffSigmaSM2} The cross section  $\sigma$  (left-top), $A_z$
		(right-top), $A_x$ (left-bottom) and $A_y$ (right-bottom) as a function of $\cos\theta$  
		of $W$ in $8$  bin in the process $e^+e^-\to W^+W^-$ for $\sqrt{s}=500$ GeV, ${\cal 
			L}=100$ fb$^{-1}$. The {\it dotted} (blue) lines corresponds to SM unpolarized values, 
		{\it solid} (red) lines corresponds to 
unpolarized {\tt aTGC} values, {\it dashed} (black) 
lines represent polarized SM values and {\it dashed-dotted} (green) lines 
represent polarized {\tt aTGC} values of 
observables. For {\tt aTGC} values only one anomalous coupling  
has been assumed non-zero and others kept at zero in each panel} 
\end{figure*}
\begin{figure*}[htb!]
	\centering
	\includegraphics[width=8.650cm]{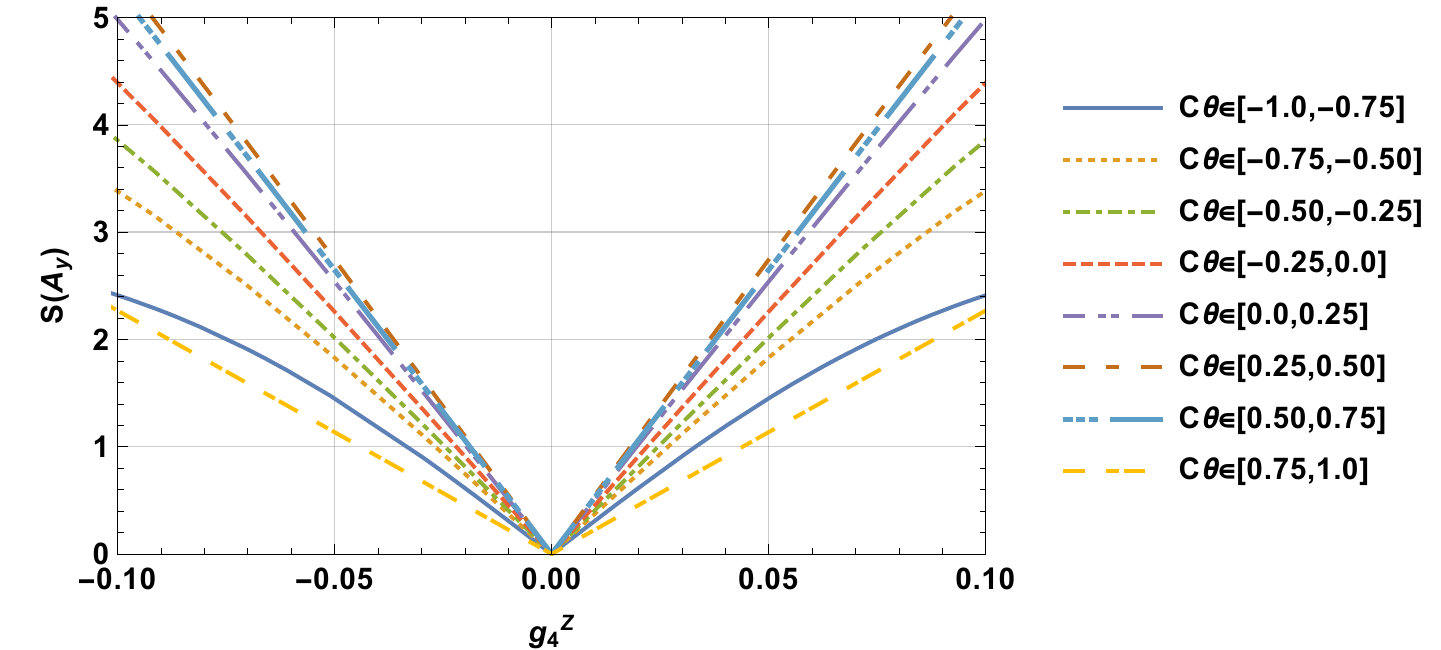}
	\includegraphics[width=8.650cm]{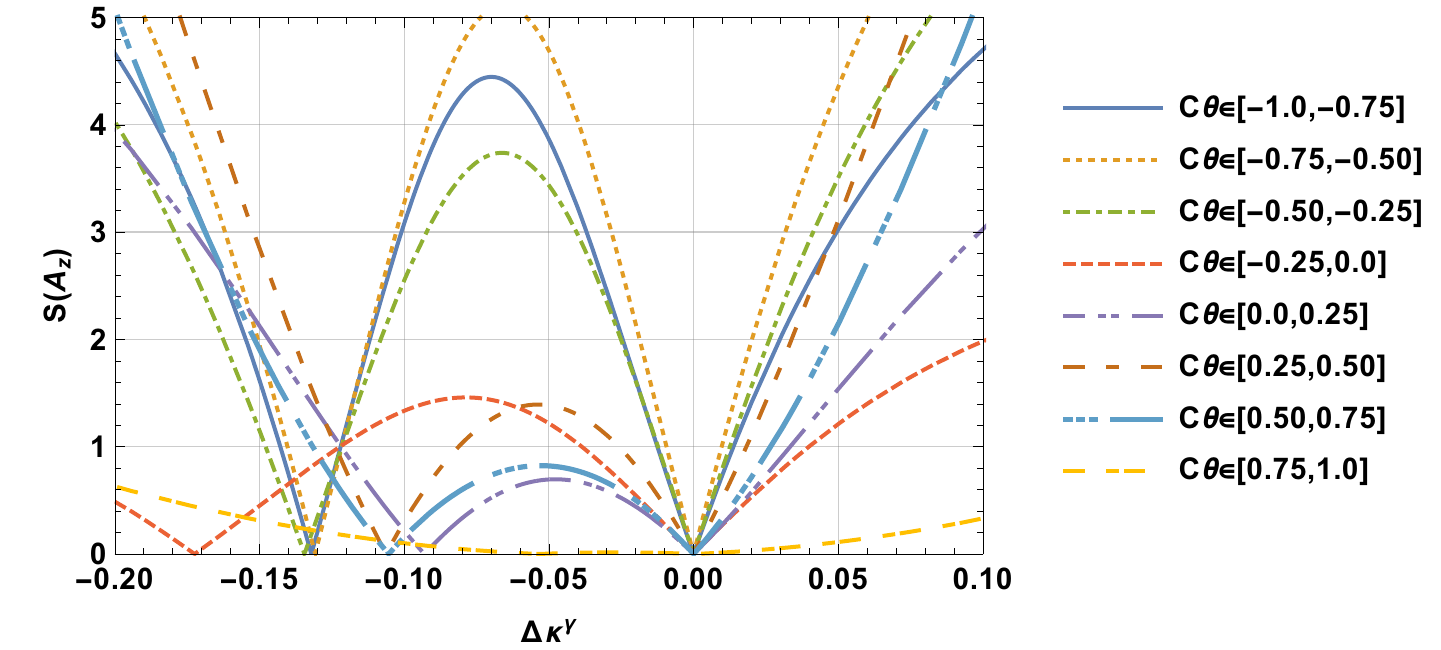}
	\caption{\label{fig:sensitivity_binned}The one parameter  sensitivity of $A_x$ 
		 on $g_4^Z$ (left) and of $A_z$ on $\Delta\kappa^\gamma$ (right) in $8$ bin
		for   $e^+e^-\to W^+W^-$ at $\sqrt{s}=500$ GeV,  ${\cal L}=100$ fb$^{-1}$, $C\theta = \cos\theta$ of $W^-$ } 
\end{figure*}
\begin{table*}[!b]\caption{\label{tab:obs-analysis-name} The list  of  
analyses performed in the present work
and  set of  observables used with different kinematical cut   to obtain 
simultaneous limits on anomalous couplings  in  $e^+e^-\to W^+W^-$  
at $\sqrt{s}=500$ GeV, ${\cal L}=100$ fb$^{-1}$ with unpolarized beams.
		 The rectangular volume of couplings at $95\%$ BCI is 
shown in the last column for each analyses (see text
		 for details)  }
	\renewcommand{\arraystretch}{1.50}
	\begin{tabular*}{\textwidth}{@{\extracolsep{\fill}}llll@{}}\hline
		Analysis name  & Set of observables & Kinematical cut on $\cos\theta$  & Volume of Limits \\\hline
	{\tt $\sigma$-ubinned} & $\sigma$ & $\cos\theta\in[-1.0,1.0]$& $4.4\times 10^{-11}$\\
		{\tt Unbinned} & 
		$\sigma$, $A_{fb}$, $A_i$
		& $\cos\theta\in[-1.0,1.0]$ & $3.1\times 10^{-12}$\\
		%$\sigma$ backward & $\sigma$ & $\cos\theta\in[-1.0,0.0]$ & $4.9 \times 10^{-11}$\\
		{\tt Backward} & $\sigma$,  $A_i$ & $\cos\theta\in[-1.0,0.0]$& $1.3 \times 10^{-13}$\\
		{\tt $\sigma$-binned} &  $\sigma$ & 
		$\cos\theta\in[\frac{m-5}{4},\frac{m-4}{4}]$, $m=1,2,3\dots, 8$
		& $5.4 \times 10^{-12}$ \\
%		{\tt Binned-best-$24$} & $\sigma$,  $A_i$ &" & $3.9 \times 10^{-14}$\\
		 {\tt Binned} & $\sigma$,  $A_i$ & "& $1.2 \times 10^{-16}$\\
%{\tt Binned-$2\sigma$ } & " & "& $9.4 \times 10^{-25}$ \\
		\hline
	\end{tabular*}
\end{table*}
Sensitivity of an observables ${\cal O}$ depending on anomalous couplings 
$\vec{f}$ with beam polarization $\eta_3$, $\xi_3$ is given by
\begin{equation}
{\cal S}({\cal O}(\vec{f},\eta_3,\xi_3))=\dfrac{|{\cal O}(\vec{f},\eta_3,\xi_3)
	-{\cal O}(\vec{0},\eta_3, \xi_3)|}{|\delta{\cal O}(\eta_3, \xi_3)|} \ \ ,
\end{equation}
where $\delta{\cal O}=\sqrt{(\delta{\cal O}_{stat.})^2+
	(\delta{\cal O}_{sys.})^2}$ is the estimated error in ${\cal O}$. 
The  error for cross-section would be,
\begin{equation}
\delta\sigma(\eta_3, \xi_3)=\sqrt{\frac{\sigma(\eta_3, \xi_3)}{{\cal L}} +
	\epsilon_\sigma^2 \sigma(\eta_3, \xi_3)^2 } \ \ ,
\end{equation}
where as the estimated error in asymmetries would be,
\begin{equation}
\delta A(\eta_3, \xi_3)=\sqrt{\frac{1-A(\eta_3, \xi_3)^2}
	{{\cal L}\sigma(\eta_3, \xi_3)} + \epsilon_A^2 } \ \ .
\end{equation}

Here ${\cal L}$ is the integrated luminosity, $\epsilon_\sigma$ and 
$\epsilon_A$ are the systematic fractional error in cross-section and 
asymmetries respectively. we take ${\cal L}=100$ fb$^{-1}$,
$\epsilon_\sigma=0.02$ and $\epsilon_A=0.01$ as a benchmark scenario for the present
analyses. 

The sensitivity of all $10$ observables have been studied on the all $14$
couplings of the Lagrangian in Eq.~\ref{eq:Lagrangian} with the chosen $\sqrt{s}$,
${\cal L}$ and systematic uncertainties. The sensitivity of all observables 
on $g_4^Z$ and $\Delta\kappa^\gamma$ are shown in Fig.~\ref{fig:sensitivity}
as representative. Being $CP$-odd (either only linear or only quadratic terms
 present) $g_4^Z$ has single patch in the confidence interval, 
while the  $\Delta\kappa^\gamma$ being  $CP$-even 
(linear and quadratic terms present), 
it has two patches in the sensitivity curve. The  $CP$-odd observable $A_y$
provides the tightest one parameter limit on $g_4^Z$. The tightest $1\sigma$ 
limit on $\Delta\kappa^\gamma$ is obtained using $A_{fb}$ while at $2\sigma$ 
level a combination of $A_{fb}$ and $A_x$ provides the tightest limit.

\begin{table*}[!t]\caption{\label{tab:Limits}List of  Posterior $95\%$ BCI
		of anomalous couplings $c_i^{\cal L}$ ($10^{-2}$) of the Lagrangian in Eq.~\ref{eq:Lagrangian} in 
		$e^+e^-\to W^+W^-$  at $\sqrt{s}=500$ GeV, ${\cal L}=100$ fb$^{-1}$ for  {\tt Unbinned} and 
		{\tt Binned} case for $5$ chosen set longitudinal beam polarizations $\eta_3$ and $\xi_3$ 
		from MCMC. The pictorial visualisation for these $95\%$ BCI of $c_i^{\cal L}$   
is shown in Fig.~\ref{fig:Limits}. The rectangular volume of couplings at $95\%$ BCI is given in the last row.
		The one parameter limits ($10^{-2}$)   at $2\sigma$ level  with unpolarized beams are given in the
		last column. The notation used here is $_{ low}^{ high}\equiv [ low,  high]$ with
		$low$ being lower limit and $high$ being upper limit}
	
	\renewcommand{\arraystretch}{1.50}
	\begin{tabular*}{\textwidth}{@{\extracolsep{\fill}}llllllllllll@{}}\hline
		$(\eta_3,\xi_3$) & \multicolumn{2}{c}{($-0.80,+0.80$)} & 	\multicolumn{2}{c}{($-0.40,+0.40$)} & 	\multicolumn{2}{c}{$(0.0,0.0)$} & 	\multicolumn{2}{c}{($+0.40,-0.40$)} & 	\multicolumn{2}{c}{($+0.80,-0.80$)} & $(0.0,0.0)$ \\\hline
		$c_i^{\cal L}$ &  {\tt Unbinned} & {\tt Binned} &  {\tt Unbinned} & {\tt Binned} &  {\tt Unbinned} & {\tt Binned} &  {\tt Unbinned} & {\tt Binned} &  {\tt Unbinned} & {\tt Binned} & $2\sigma$ limit\\\hline
		$\Delta g_1^{\gamma}$			& $_{-38}^{+48}$	& $_{-25}^{+17} $	& $_{-30}^{+34} $	& $_{-17}^{+11} $	& $_{-15}^{+18} $	& $_{-8.9}^{+5.9} $	& $_{-12}^{+13} $	& $_{-7.5}^{+2.8} $	& $_{-39}^{+28} $	& $_{-10}^{+14} $	& $_{-1.5}^{+1.4} $\\
		$g_4^{\gamma}$ 					& $_{-46}^{+47}$	& $_{-23}^{+23} $	& $_{-28}^{+26} $	& $_{-11}^{+12} $	& $_{-13}^{+13} $	& $_{-6.2}^{+6.3} $	& $_{-8.3}^{+8.3} $	& $_{-4.4}^{+4.5} $	& $_{-15}^{+16} $	& $_{-9.1}^{+9.2} $	& $_{-2.1}^{+2.1} $\\ 
		$g_5^{\gamma}$ 					& $_{-47}^{+44}$	& $_{-25}^{+23} $	& $_{-30}^{+27} $	& $_{-12}^{+12} $	& $_{-15}^{+11} $	& $_{-6.5}^{+6.2} $	& $_{-9.3}^{+8.0} $	& $_{-4.3}^{+3.4} $	& $_{-7.0}^{+28} $	& $_{-8.3}^{+7.7} $	& $_{-2.2}^{+2.0} $\\
		
		$\lambda^{\gamma}$& $_{-22}^{+18}$			& $_{-9.3}^{+9.0} $	& $_{-7.3}^{+6.5}$	& $_{-2.8}^{+2.8} $	& $_{-4.0}^{+3.2}$	& $_{-1.9}^{+1.5}$	& $_{-3.4}^{+2.5} $	& $_{-1.7}^{+1.1}$	& $_{-4.6}^{+5.3} $	& $_{-4.1}^{+2.5} $	& $_{-1.2}^{+0.85} $\\
		$\widetilde{\lambda^{\gamma}}$  & $_{-23}^{+18}$	& $_{-8.8}^{+8.5} $	& $_{-6.9}^{+6.9} $	& $_{-3.0}^{+2.8} $	& $_{-3.6}^{+3.6} $	& $_{-1.6}^{+1.6}$	& $_{-2.9}^{+2.9}$	& $_{-1.4}^{+1.4}$	& $_{-5.0}^{+5.3} $	& $_{-2.7}^{+3.1} $	& $_{-1.2}^{+1.2} $\\
		$\Delta\kappa^{\gamma}$& $_{-28}^{+17}$	& $_{-11}^{+8} $	& $_{-15.2}^{+4.1} $	& $_{-7.6}^{+3.3} $	& $_{-10.7}^{-0.3}$	& $_{-5.7}^{+1.0} $	& $_{-11.2}^{-0.8}$	& $_{-5.8}^{+0.7} $	& $_{-24}^{+13} $	& $_{-14.4}^{+2.9} $	& $_{-0.37}^{+0.36} $\\ 
		$\widetilde{\kappa^{\gamma}}$ 	& $_{-50}^{+48} $	& $_{-23}^{+29}$	& $_{-27}^{+27}$	& $_{-12}^{+12}$	& $_{-14}^{+14}$	& $_{-6.4}^{+6.1} $	& $_{-10}^{+9.9} $	& $_{-4.7}^{+4.7}$	& $_{-21}^{+21} $	& $ _{-8.1}^{+9.3}$	& $_{-0.26}^{+0.26} $\\ 
		$\Delta g_1^Z$ 					& $_{-33}^{+37}$			& $ _{-13}^{+20} $	& $_{-21}^{+29}$	& $_{-9.0}^{+13}$	& $_{-9.0}^{+19}$	& $_{-4.0}^{+7.7} $	& $_{-7.0}^{+16} $	& $_{-3.7}^{+5.8} $	& $_{-33}^{+32} $	& $_{-10}^{+19} $	& $_{-1.4}^{+1.4} $\\ 
		$g_4^Z$ 						& $_{-38}^{+38}$			& $_{-18}^{+18}$	& $_{-21}^{+21}$	& $_{-9.3}^{+8.4} $	& $_{-10}^{+10}$	& $_{-5.0}^{+5.1} $	& $_{-7.3}^{+7.2} $	& $_{-4.2}^{+4.0} $	& $_{-15}^{+15}$	& $_{-8.9}^{+9.1} $	& $_{-1.5}^{+1.5} $\\ 
		$g_5^Z$ 						& $_{-39}^{+34} $	& $_{-19}^{+18} $	& $_{-24}^{+20}$	& $_{-9.9}^{+8.5}$	& $_{-13}^{+9.0}$	& $_{-4.9}^{+4.9} $	& $_{-8.1}^{+6.8} $	& $_{-3.7}^{+3.0} $	& $_{-8.5}^{+26.5}$	& $_{-8.2}^{+7.5} $	& $_{-1.3}^{+1.3} $\\ 
		$\lambda^Z$ 					& $_{-14}^{+17} $	& $_{-7.5}^{+6.9} $	& $_{-5.8}^{+5.2}$	& $_{-2.4}^{+2.0}$	& $_{-3.5}^{+2.7}$	& $_{-1.5}^{+1.1}$	& $_{-3.0}^{+2.3} $	& $_{-1.6}^{+0.8}$	& $ _{-4.4}^{+5.2}$	& $_{-3.8}^{+2.3} $	& $_{-0.6}^{+0.6} $\\ 
		$\widetilde{\lambda^Z}$ 		& $_{-14}^{+17} $	& $_{-6.6}^{+6.8} $	& $_{-5.5}^{+5.5}$	& $_{-2.3}^{+2.3}$	& $_{-3.1}^{+3.1}$	& $_{-1.3}^{+1.3}$	& $_{-2.7}^{+2.7}$	& $_{-1.2}^{+1.2}$	& $_{-4.4}^{+5.0} $	& $_{-2.6}^{+3.0} $	& $_{-0.6}^{+0.6} $\\ 
		$\Delta\kappa^Z$ 				& $_{-21}^{+15}$			& $_{-6.5}^{+9.7}  $& $_{-9.6}^{+5.6}$	& $_{-3.4}^{+5.2} $	& $_{-6.5}^{+2.5}$	& $_{-1.7}^{+3.6} $	& $_{-6.4}^{+2.3} $	& $_{-2.0}^{+3.5} $	& $_{-19}^{+14} $	& $_{-10.6}^{+5.7} $	& $_{-0.5}^{+0.48} $\\
		$\widetilde{\kappa^Z}$ 			& $_{-32}^{+41} $	& $_{-22}^{+18}$	& $_{-22}^{+22}$	& $_{-8.8}^{+9.7}$	& $_{-11}^{+11}$	& $_{-4.8}^{+5.3} $	& $_{-9.0}^{+8.9} $	& $_{-4.3}^{+4.2} $	& $_{-22}^{+21} $	& $_{-8.0}^{+8.4} $	& $_{-1.6}^{+1.6} $\\ 
		$Volume$ &$_{5.1\times 10^{-4}}$& $_{1.6\times 10^{-8}}$ &$_{2.9\times  10^{-8}}$&$_{3.0\times 10^{-13}}$ &$_{3.0\times  10^{-12}}$&$_{1.0\times 10^{-16}}$ &$_{1.1\times  10^{-13}}$&$_{5.0\times 10^{-18}}$ &$_{1.0\times  10^{-8}}$&$_{6.0\times 10^{-13}}$&$_{9.4 \times 10^{-25}}$   \\
		\hline
	\end{tabular*}
\end{table*}
%%%%%%%%%%%%%%%%%%%%%%%%%%%%%%%%%%%%%%%%%%%%%%%%%%%%%%%%%%%%%%%%%%%%%%%%%%%%%%%%%%%%%

Here, we have a total of $14$ different anomalous couplings to measure, while 
we only have $10$ observables. A certain combination of large couplings may 
mimic the SM within the statistical errors. To avoid these we need more number 
of observables to be included in the analysis. To this end we divide $\cos\theta$
(production angle of $W$) into eight bins and calculate the cross section and
polarization asymmetries in all of them.
The cross section and  the polarization asymmetries $A_z$, $A_{xz}$ and $A_y$ as 
a function of $\cos\theta$ are shown in Fig.~\ref{fig:diffSigmaSM2} for the SM and {\tt aTGC}
for both polarized and unpolarized beams.
The SM values for unpolarized case is shown in {\it dotted} (blue),
SM with polarization of $(\eta_3,\xi_3)=(+0.8,-0.8)$ is shown in {\it dashed} (black) lines.
The {\it solid} (red) lines corresponds to unpolarized {\tt aTGC} values while   {\it dashed-dotted}
 (green) lines represent polarized {\tt aTGC} values of observables. 
For the cross section (left-top panel) we take $\Delta g_1^\gamma$  to be $0.1$  and all other 
couplings to zero for both the polarized and unpolarized case.
  We see that the fractional deviation from the SM values is larger in the the
most backward bin $\cos\theta\in(-1.0 , -0.75)$ and gradually reduces in the
forward direction.  The deviation is even larger in case
  of beam polarization.
 Sensitivity of cross section on  $\Delta g_1^\gamma$  is thus expected to be high in the most 
 backward bin. In case of asymmetries $A_z$ (right-top panel), $A_{xz}$ (left-bottom panel) 
 and $A_y$ (right-bottom panel) the {\tt aTGC} is assumed to be $\Delta\kappa^Z=0.05$, 
 $\lambda^Z=0.05$  and $g_4^Z=0.05$, respectively, while others are kept at 
zero. The change in the asymmetries due to {\tt aTGC} is larger in the backward
bin for both the beam polarizations. 
%For $A_z$ and $A_y$ in the right panel we see the asymmetry  is positive in some bins,
%  while negative in some other bins  enhancing the values (modulus) in each bin.
The value of asymmetries in each bin are comparable to the
 total values. The asymmetries may not have highest sensitivity in the most backward bin but 
 in other bin. So we can not ignore observables in any bin, we will
 use all $9$ observables ($A_{fb}$ excluded) in $8$ bin totalling $72$ observables in our
 analysis. 
 %We consider only $8$ bins and do not increase them for simplicity to handle
 %the numerical calculations and also because increase of bin numbers reduces the
 % number of events and hence increment of statistical error.
  
One parameter sensitivity of the set of $9$ observables in $8$ bin has
been studied. We show sensitivity of $A_y$  on $g_4^Z$ and of $A_z$ on $\Delta\kappa^\gamma$ 
in the $8$ bin in Fig.~\ref{fig:sensitivity_binned}. We can see the tightest limits based on 
sensitivity  got much tighter (tighten by a factor of $2$, roughly) compared 
to the unbin case in Fig.~\ref{fig:sensitivity}. Thus we expect simultaneous limits
on all the couplings to be tighter when observables get binned.

We perform a set of MCMC analyses with different set of observables for different kinematical
cuts with unpolarized beams to understand their roles in providing limits on
the anomalous couplings. These analyses are listed in 
Table~\ref{tab:obs-analysis-name}. 
The corresponding $14$ dimensional rectangular volume made out of $95\%$ Bayesian confidence
interval (BCI) on the anomalous couplings are also listed in
Table~\ref{tab:obs-analysis-name} in the last column. 
The simplest analysis would be to consider only cross section in the full $\cos\theta$ domain
and perform MCMC analysis which is named as  {\tt $\sigma$-ubinned}. The typical
$95\%$ limits on the parameters range from $\sim \pm0.04$ to $\pm0.25$ giving the volume of 
limits to be $4.4\times 10^{-11}$. As we have polarizations asymmetries, the straight
forward analysis would be to consider all observables  
for full domain of $\cos\theta$. This analysis is named  {\tt Unbinned} where  limits on 
anomalous couplings get constrained better reducing the volume of limits by a factor of $10$
compared to the {\tt $\sigma$-ubinned}. Motivated by $\cos\theta$ dependence of the observables (in Fig.~\ref{fig:diffSigmaSM2}) we perform the analysis {\tt Backward} 
using all the observables ($A_{fb}$ is no longer observables here) in the backward direction, i.e. $\cos\theta\in[-1.0,0.0]$. In the {\tt Backward}  case the limits get improved further, the
 volume of limit reduces by a factor of $10$ compared to the {\tt Unbinned} case. 
 To see how binning improve the limits we perform a analysis named {\tt $\sigma$-binned} 
 using only cross section in $8$ bin. We see the analysis {\tt $\sigma$-binned} is better 
 than {\tt $\sigma$-unbinned} and comparable to the analysis {\tt Unbinned} but not better 
 than the analysis {\tt Backward}. 
 %Next in the analysis named {\tt Binned-best-$24$} we consider a minimal set of observables  
 %consisting best $24$  out of $72$ observables after  binning. We comprise the minimal set choosing 
 %only the observables providing tightest one parameters limits on the couplings  best on sensitivity. The analysis {\tt Binned-best-$24$} turn out be better than all previous case. 
The most natural and complete analysis would be to  consider all the  observables after binning.
The analysis is named as  {\tt Binned} which has  limits  much better than any analysis. 
We also calculate one parameters limit on all the $14$ couplings at $2\sigma$
sensitivity ($\chi^2=4$) considering all the binned observables for 
unpolarized beams. The one parameter limits are presented in Table~\ref{tab:Limits} in the last 
column for comparison.
Although the  {\tt Binned} analysis provides best limits it is natural to perform
{\tt Unbinned} analysis also for comparison. We perform these two analysis 
for a set of beam polarization in the next subsection to obtain better limits
that the one provided by unpolarized beams.
%%%%%%%%%%%%%%%%%%%%%%%%%%%%%%%%%%%%%%%%%%
  \begin{figure*}[!t]
 	\centering
 	\includegraphics[width=8.6cm]{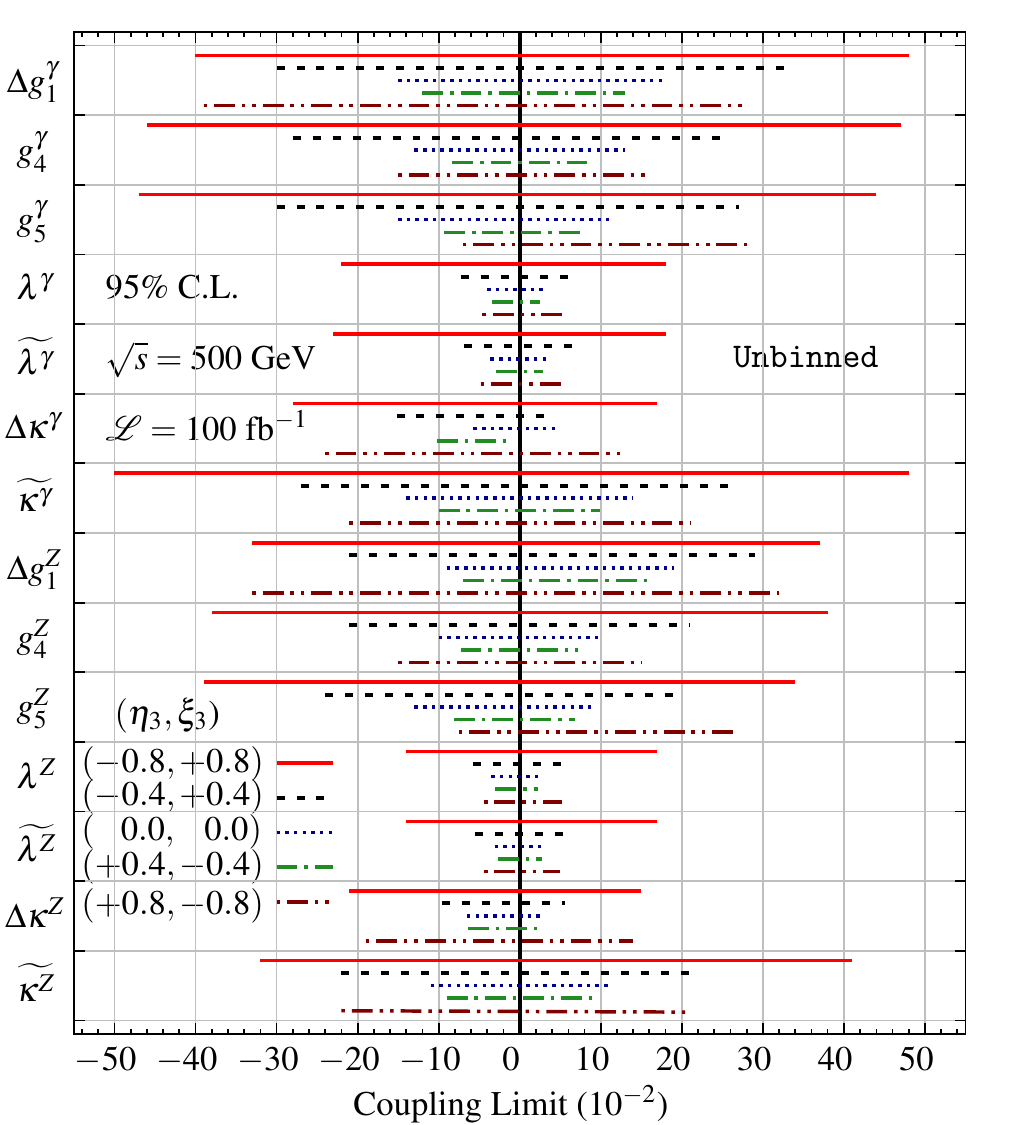}
 	\includegraphics[width=8.6cm]{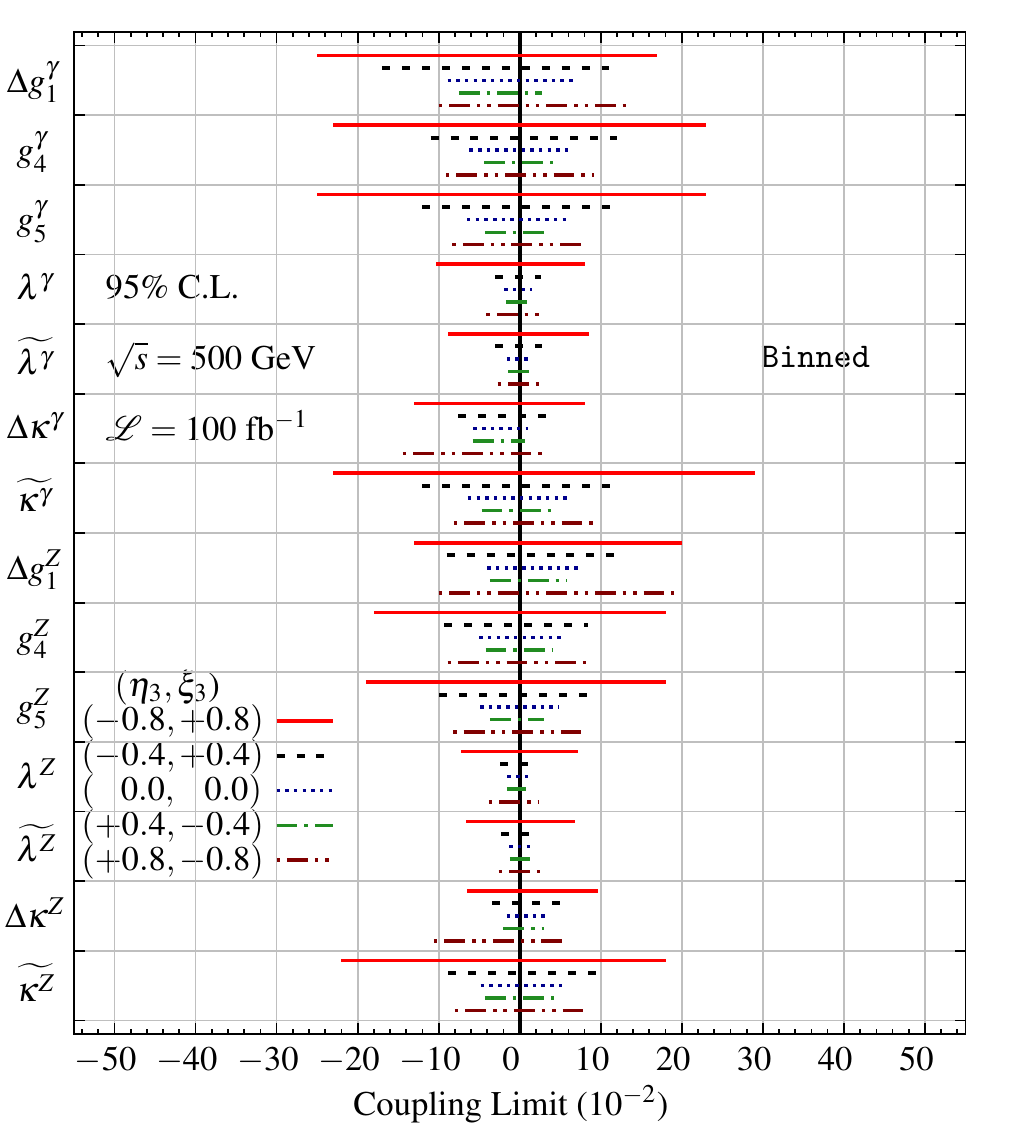}
 	\caption{\label{fig:Limits} The pictorial visualisation of $95\%$ BCI of anomalous couplings $c_i^{\cal L}$     
 		in $e^+e^-\to W^+W^-$ for $\sqrt{s}=500$ GeV, ${\cal L}=100$ 
 		fb$^{-1}$ for  {\tt Unbinned} case  (left panel) and {\tt Binned} case (right panel). The numerical values of the 
 		limits  can be read of  in Table~\ref{tab:Limits}} 
 \end{figure*}

 \begin{table*}[!t]
 	\caption{\label{tab:Limits_Operator}The list of  posterior   $95\%$ BCI
 		of anomalous couplings $c_i^{\cal O}$ (TeV$^{-2}$)  of effective operators  in Eq.~\ref{eq:OW} 
in $e^+e^-\to W^+W^-$  at $\sqrt{s}=500$ GeV, ${\cal L}=100$ fb$^{-1}$ for  {\tt Unbinned} and 
 		{\tt Binned} case for $5$ chosen set of longitudinal beam polarizations $\eta_3$ and $\xi_3$ from MCMC.
 		The pictorial visualisation for these $95\%$ BCI of $c_i^{\cal O}$ is shown in   Fig.~\ref{fig:Limits-Gauged}.
 		The rectangular volume (TeV$^{-10}$) of couplings at $95\%$ BCI is given in the last row. Rest details are same 
 		as in Table~\ref{tab:Limits} }
 	\renewcommand{\arraystretch}{1.50}
 	\begin{tabular*}{\textwidth}{@{\extracolsep{\fill}}llllllllllll@{}}\hline
 		$(\eta_3,\xi_3$) & \multicolumn{2}{c}{($-0.80,+0.80$)} & 	
 		\multicolumn{2}{c}{($-0.40,+0.40$)} & 	\multicolumn{2}{c}{$(0.0,0.0)$} & 	
 		\multicolumn{2}{c}{($+0.40,-0.40$)} & 	\multicolumn{2}{c}{($+0.80,-0.80$)}&  $(0.0,0.0)$\\\hline
 		$c_i^{\cal O}$                                 &  {\tt Unbinned} & {\tt Binned} &  {\tt Unbinned} & {\tt Binned} &  {\tt Unbinned} & {\tt Binned} &  {\tt Unbinned} & {\tt Binned} &  {\tt Unbinned} & {\tt Binned} & $2\sigma$ limit\\\hline
 		$\frac{c_{WWW}}{\Lambda^2}$             &$_{-5.7}^{+6.2}$ &$_{-0.8}^{+0.9}$  &$_{-4.8}^{+6.6}$ &$_{-1.2}^{+1.1}$ &$_{-4.9}^{+6.7} $ &$_{-2.0}^{+1.3}$ &$_{-7.5}^{+7.4}$  &$_{-3.5}^{+1.7}$ &$_{-13.7}^{+8.7}$ &$_{-10.5}^{+4.5}$ &$_{-1.0}^{+1.0}$ \\
 		$\frac{c_{W}}{\Lambda^2}$               &$_{-12.5}^{+35.9}$ &$_{-3.9}^{+3.8}$  &$_{-8.5}^{+9.8}$ &$_{-2.2}^{+5.2}$ & $_{-6.6}^{+5.7} $ &$_{-1.4}^{+5.0}$  &$_{-7.4}^{+6.2}$ &$_{-1.3}^{+4.9}$ &$_{-8.0}^{+5.4}$ &$_{-2.1}^{+16.7}$&$_{-0.6}^{+0.6}$\\
 		$\frac{c_{B}}{\Lambda^2}$               &$_{-18.1}^{+46.0}$  &$_{-15.9}^{+15.3}$  &$_{-53.4}^{+20.9}$ &$_{-21.5}^{+7.6}$ & $_{-28.8}^{+6.7} $ &$_{-23.7}^{+2.8}$  &$_{-21.1}^{+0.6}$ &$_{-21.1}^{+0.9}$ &$_{-3.1}^{+1.0}$ &$_{-16.2}^{+0.6}$&$_{-1.3}^{+1.3}$\\
 		$ \frac{c_{\widetilde{WWW}}}{\Lambda^2}$&$_{-6.3}^{+6.3}$  &$_{-0.8}^{+0.8}$ &$_{-6.2}^{+6.1}$ &$_{-1.0}^{+1.0}$ &$_{-5.8}^{+5.8} $ &$_{-1.6}^{+1.6}$  &$_{-6.8}^{+6.9}$ &$_{-2.5}^{+2.5}$ &$_{-11.7}^{+10.2}$ &$_{-5.8}^{+5.7}$&$_{-1.1}^{+1.1}$\\
 		$ \frac{c_{\widetilde{W}}}{\Lambda^2}$  &$_{-5.6}^{+5.6}$  &$_{-9.8}^{+10.1}$ &$_{-49.1}^{+48.6}$ & $_{-11.2}^{+11.5}$&$_{-40.5}^{+40.3} $ &$_{-13.5}^{+13.0}$  &$_{-29.0}^{+29.5}$ &$_{-6.8}^{+6.8}$ &$_{-6.7}^{+7.4}$ &$_{-2.5}^{+2.5}$&$_{-12.0}^{+12.0}$\\
 		$Volume$ &$_{1.8 \times 10^{8}}$&$_{1.5 \times 10^{4}}$&$_{1.9 \times 10^{7}}$&$_{2.3 \times 10^{4}}$&$_{4.8 \times 10^{6}}$&$_{4.9 \times 10^{5}}$&$_{3.6 \times 10^{6}}$&$_{5.0 \times 10^{4}}$&$_{3.9 \times 10^{5}}$&$_{2.8 \times 10^{5}}$&$_{3.2\times 10^2}$\\
 		
 		\hline
 	\end{tabular*}
 \end{table*}
 
 \begin{table*}[!t]\caption{\label{tab:Limits-gauged}The list  of posterior   
$95\%$ BCI of anomalous couplings $c_i^{{\cal L}_g}$ ($10^{-2}$) translated  
from  $c_i^{\cal O}$ (using Eq.~\ref{eq:Operator-to-Lagrangian})
 		in  $SU(2)\times U(1)$ gauge in $e^+e^-\to W^+W^-$  at 
$\sqrt{s}=500$ GeV, ${\cal L}=100$ fb$^{-1}$. Other details are same as in Table~\ref{tab:Limits}}
 	\renewcommand{\arraystretch}{1.50}
 	\begin{tabular*}{\textwidth}{@{\extracolsep{\fill}}llllllllllll@{}}\hline
 		$(\eta_3,\xi_3$) & \multicolumn{2}{c}{($-0.80,+0.80$)} & 	\multicolumn{2}{c}{($-0.40,+0.40$)} & 	\multicolumn{2}{c}{$(0.0,0.0)$} & 	\multicolumn{2}{c}{($+0.40,-0.40$)} & 	\multicolumn{2}{c}{($+0.80,-0.80$)}&  $(0.0,0.0)$ \\\hline
 		$c_i^{{\cal L}_g}$               &  {\tt Unbinned} & {\tt Binned} &  {\tt Unbinned} & {\tt Binned} &  {\tt Unbinned} & {\tt Binned} &  {\tt Unbinned} & {\tt Binned} &  {\tt Unbinned} & {\tt Binned}& $2\sigma$ limit \\\hline
 		$\lambda^V$                  &$_{-2.3}^{+2.5}$ &$_{-0.3}^{+0.4}$ &$_{-1.9}^{+2.7}$ &$_{-0.5}^{+0.4}$ &$_{-2.0}^{+2.7}$ &$_{-0.8}^{+0.5}$ &$_{-3.1}^{+3.0}$ &$_{-1.4}^{+0.6}$&$_{-5.6}^{+3.5}$ &$_{-4.3}^{+1.8}$&$_{-0.4}^{+0.4}$ \\
 		$\widetilde{\lambda^V}$      &$_{-2.5}^{+2.6}$ &$_{-0.3}^{+0.3}$ &$_{-2.5}^{+2.5}$ &$_{-0.4}^{+0.4}$ &$_{-2.3}^{+2.3}$ &$_{-0.6}^{+0.6}$ &$_{-2.8}^{+2.8}$ &$_{-1.0}^{+1.0}$&$_{-4.8}^{+4.2}$ &$_{-2.3}^{+2.3}$ &$_{-0.4}^{+0.4}$\\
 		$\Delta\kappa^{\gamma}$      &$_{-47.0}^{+11.0}$ &$_{-3.9}^{+3.7}$ &$_{-14.9}^{+4.6}$ &$_{-5.3}^{+1.7}$ &$_{-8.7}^{+1.5}$ &$_{-6.3}^{+0.5}$ &$_{-7.6}^{+0.5}$ &$_{-6.3}^{+0.5}$&$_{-3.3}^{+1.9}$ &$_{-1.0}^{+0.7}$&$_{-0.6}^{+0.6}$ \\
 		$\widetilde{\kappa^{\gamma}}$&$_{-18.3}^{+18.2}$ &$_{-3.1}^{+3.2}$ &$_{-15.8}^{+15.7}$ &$_{-3.6}^{+3.7}$ &$_{-13.0}^{+13.0}$ &$_{-4.3}^{+4.2}$ &$_{-9.3}^{+9.5}$ &$_{-2.1}^{+2.2}$ &$_{-2.1}^{+2.4}$ &$_{-0.8}^{+0.8}$&$_{-4.2}^{+4.2}$\\
 		$\Delta g_1^Z$               &$_{-5.2}^{+14.9}$   &$_{-1.6}^{+1.5}$ &$_{-3.5}^{+4.1}$ &$_{-0.9}^{+2.1}$ &$_{-2.7}^{+2.3}$ &$_{-0.6}^{+2.1}$ &$_{-3.0}^{+2.5}$ &$_{-0.5}^{+2.0}$&$_{-3.3}^{+2.2}$ &$_{-0.9}^{+6.9}$ &$_{-0.25}^{+0.25}$\\
 		$\Delta\kappa^Z$             &$_{-8.2}^{+28.3}$ &$_{-2.6}^{+2.7}$ &$_{-4.4}^{+7.7}$ &$_{-1.4}^{+3.6}$ &$_{-2.3}^{+4.0}$ &$_{-0.7}^{+3.6}$ &$_{-1.8}^{+3.2}$ &$_{-0.5}^{+3.4}$&$_{-2.3}^{+1.6}$ &$_{-0.6}^{+6.7}$&$_{-0.07}^{+0.07}$ \\
 		$\widetilde{\kappa^Z}$       &$_{-5.3}^{+5.3}$ &$_{-0.9}^{+0.9}$ &$_{-4.5}^{+4.5}$ &$_{-1.0}^{+1.0}$ &$_{-3.7}^{+3.7}$ &$_{-1.2}^{+1.2}$ &$_{-2.7}^{+2.7}$ &$_{-0.6}^{+0.6}$&$_{-0.7}^{+0.6}$ &$_{-0.2}^{+0.2}$ &$_{-1.1}^{+1.1}$\\
 		$Volume$ &$_{4.1 \times 10^{-6}}$&$_{8.3 \times 10^{-12}}$&$_{1.2 \times 10^{-7}}$&$_{1.4 \times 10^{-11}}$&$_{ 1.4\times 10^{-8}}$&$_{3.1 \times 10^{-11}}$&$_{ 8.4\times 10^{-9}}$&$_{1.7 \times 10^{-11}}$&$_{5.9 \times 10^{-10}}$&$_{2.3 \times 10^{-11}}$&$_{1.0 \times 10^{-14}}$\\
 		
 		\hline
 	\end{tabular*}
 \end{table*}
 \begin{figure*}[!h]
 	\centering
 	\includegraphics[width=7.5cm]{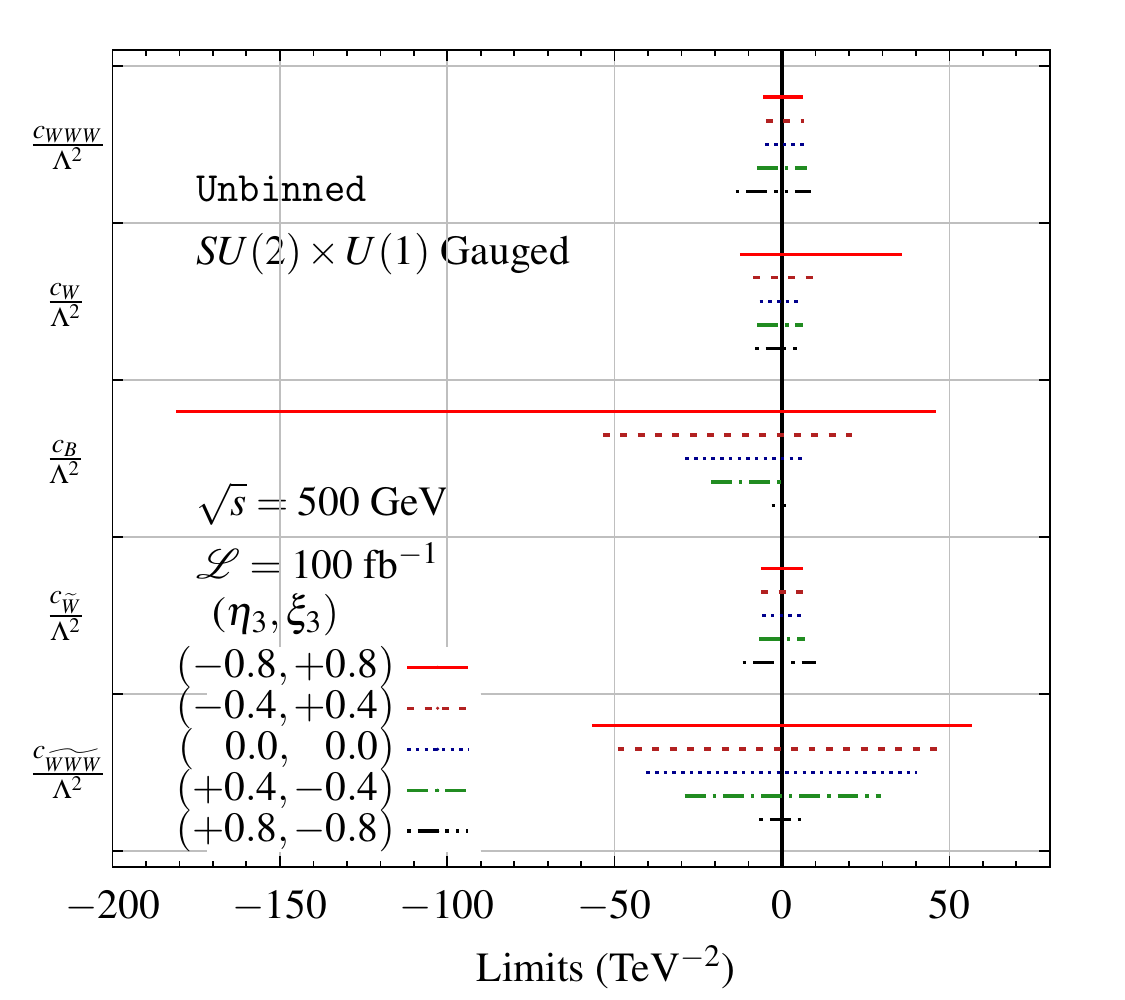}
 	\includegraphics[width=7.5cm]{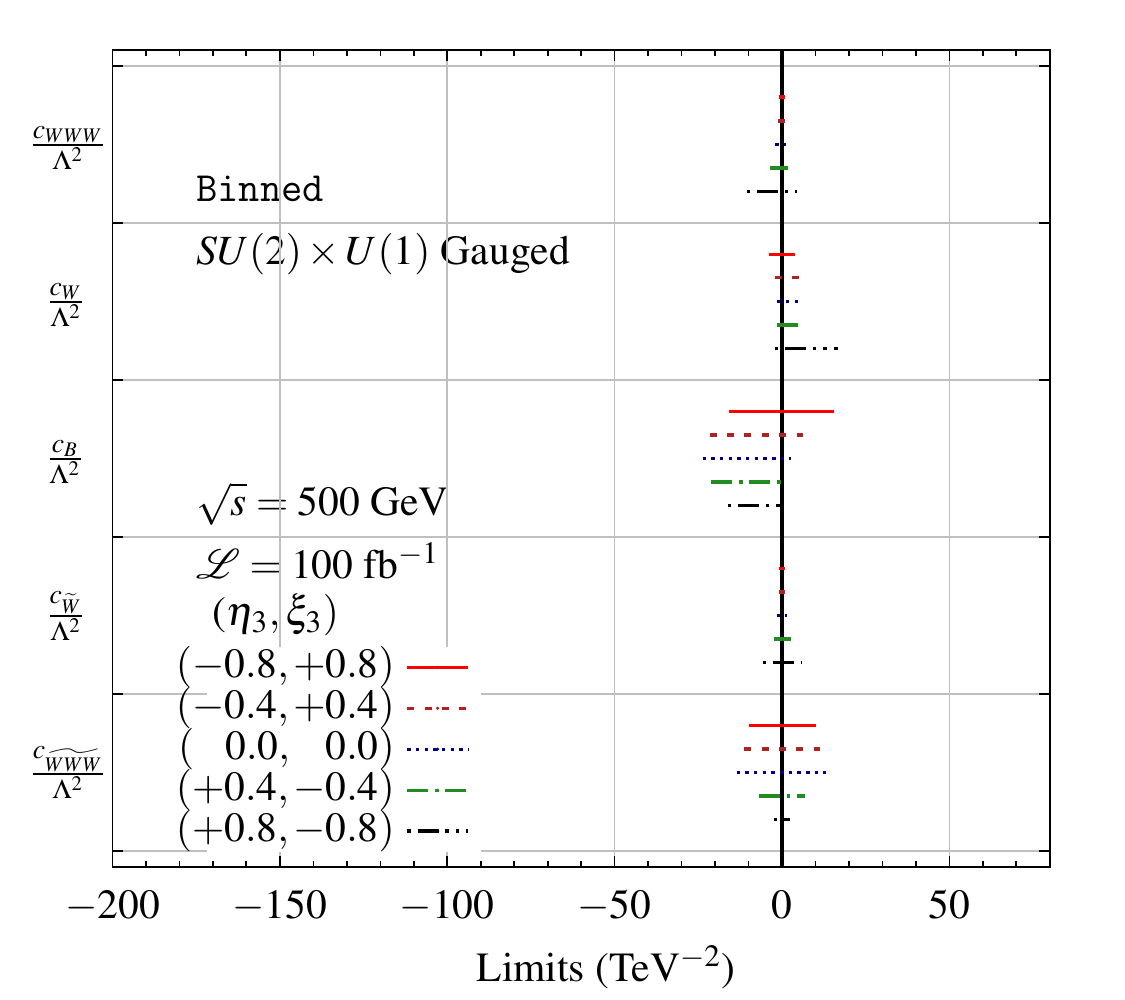}\\
 	\includegraphics[width=7.5cm]{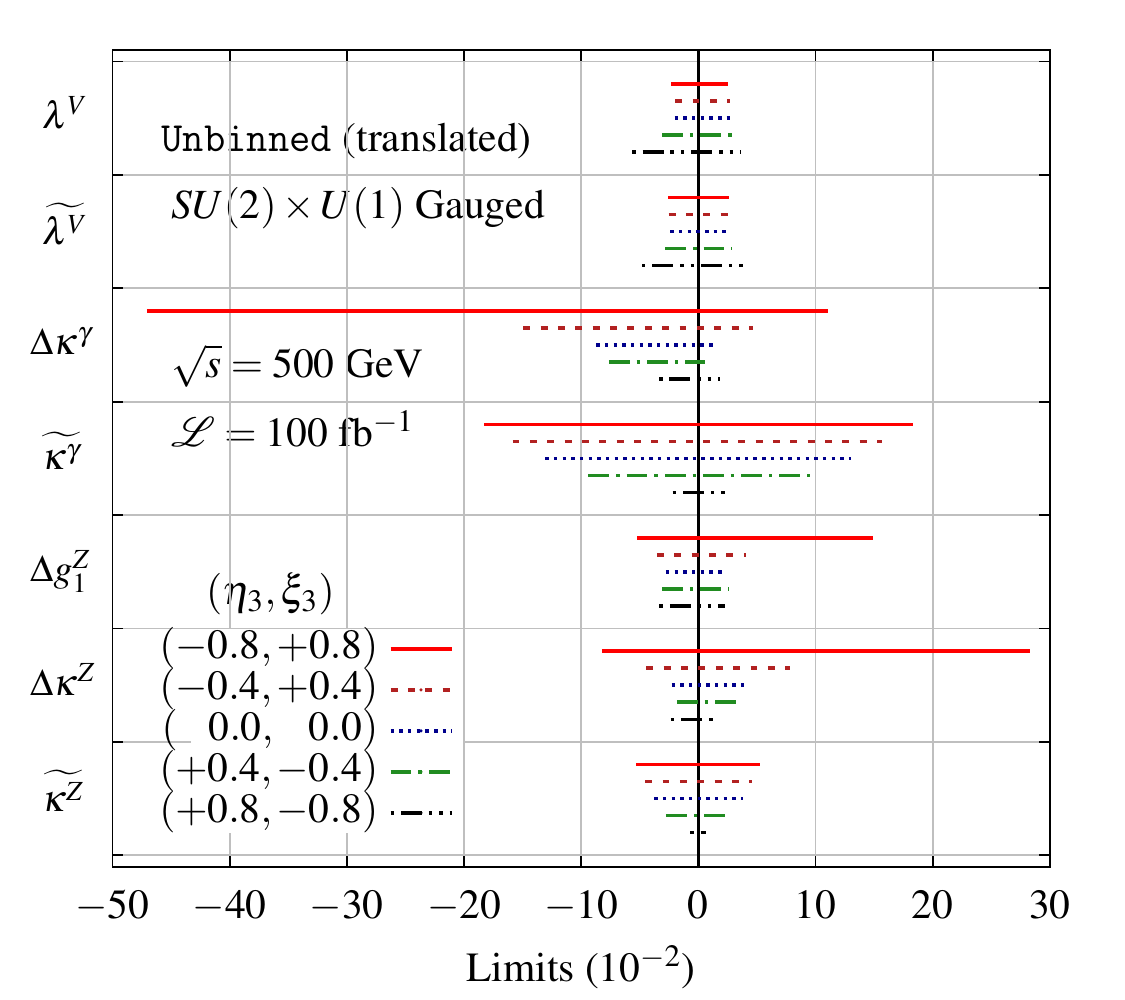}
 	\includegraphics[width=7.5cm]{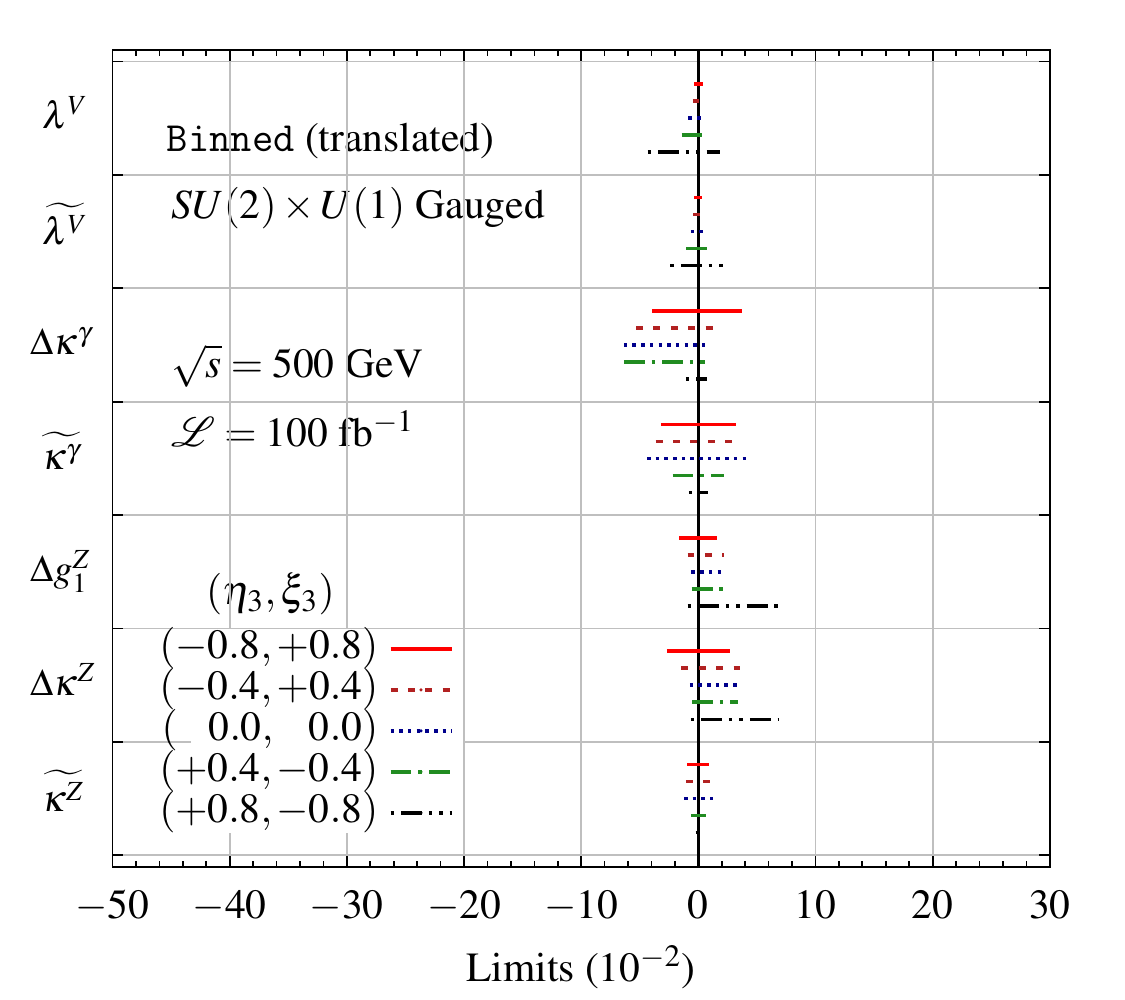}
 	\caption{\label{fig:Limits-Gauged}
 		The pictorial visualisation of simultaneous limits on anomalous couplings $c_i^{\cal O}$ of 
 		effective operators (in Eq.~\ref{eq:opertaors5}) on top row and  translated limits on $c_i^{{\cal L}_g}$ 
 		of the Lagrangian (using Eq.~\ref{eq:Operator-to-Lagrangian}) on bottom 
row in $SU(2)\times U(1)$ gauge  at $95\%$ C.L   in 
 		$e^+e^-\to W^+W^-$ for $\sqrt{s}=500$ GeV, ${\cal L}=100$ fb$^{-1}$ for 
 {\tt Unbinned} case (left column) and {\tt Binned} case (right column). The numerical 
 		values of the limits can be read of  in Tables~\ref{tab:Limits_Operator}
 		and~\ref{tab:Limits-gauged} } 
 \end{figure*}

%%%%%%%%%%%%%%%%%%%%%%%%%%%%%%%%%%%%%%%%%%%%%%%%%%%%%%%%%%%%%%%%%%%%%%%%%%%%%%%%%%%%%%%%%%%%%%%%%%%%
\subsection{Limits on the  Lagrangian couplings $c_i^{\cal L}$}\label{sec:3.2} 
We estimate simultaneous limit on all $14$ (independent) anomalous couplings  of
the Lagrangian in Eq.~\ref{eq:Lagrangian}  using MCMC method 
for {\tt Unbinned} and {\tt Binned} case.
The $95\%$ simultaneous limits on anomalous couplings  are shown in Table~\ref{tab:Limits}
 for five different set of  chosen beam polarization 
($\eta_3,\xi_3$)  namely $(-0.8,+0.8)$, $(-0.4,+0.4)$, 
$(0.0,+0.0)$, $(+0.4,-0.4)$ and 
$(+0.8,-0.8)$, which are along the cross-diagonal of  $\eta_3,\xi_3$ plane. The choice of beam
polarizations is made motivated by the result in Ref.~\cite{Rahaman:2017qql} where best choice
of beam polarization are along the  $\eta_3=-\xi_3$ line.
While presenting limits the notation used is following
$$_{ low}^{ high}\equiv [ low,  high].$$
We observe that the limits on anomalous couplings are tightest for the polarization 
$(+0.4,-0.4)$ in both {\tt Unbinned}   and   {\tt Binned} case  and in each case of polarization
 the  {\tt Binned} limits are roughly twice better than the  {\tt Unbinned} limits on average. 
We estimate simultaneous limit on couplings on several other polarization point 
along $\eta_3=-\xi_3$ direction and find the $(+0.4,-0.4)$ polarization to be the best 
to provide tightest limit. To check that $(+0.4,-0.4)$ to be best polarization
 we make a finer grid  with $9$ polarization
points around it as $$(\eta_3,\xi_3)=(\{+0.35,+0.40,+0.45\},\{-0.35,-0.40,-0.45\})$$
and find simultaneous limits on them.
We find that the limits on the couplings in these points are roughly same 
with slight variation and $(+0.4,-0.4)$ is best among them in both {\tt Unbinned}   and   
{\tt Binned} case. The lowest row in the Table~\ref{tab:Limits} shows the volume of 
the rectangular box that is formed by the limits of the couplings at the $95\%$ BCI. 
The volume is smallest for {\tt Binned}
with polarization $(+0.4,-0.4)$ as discussed above. A pictorial representation
of the limits on couplings given  in  Table~\ref{tab:Limits} is shown in
Fig.~\ref{fig:Limits} for the easy comparisons.

%%%%%%%%%%%%%%%%%%%%%%%%%%%%%%%%%%%%%%%%%%%%%%%%%%%%%%%%%%%%%%%%%%%%%%%%%%%%%%%%%%%%%%%%%%%%%%%%

\subsection{Limits on operator couplings $c_i^{\cal O}$ and their translation to   couplings $c_i^{{\cal L}_g}$ in  $ SU(2)\times U(1)$ gauge}\label{sec:3.3}

We also study the anomalous charge gauge boson couplings in $e^+e^-\to W^+W^-$ in the
framework of effective higher dimensional operator in Eq.~\ref{eq:opertaors5} in the
 $SU(2)\times U(1)$ gauge.
Similar to the case of effective vertex formalism, we calculate one parameter limit on
the  couplings at $2\sigma$ sensitivity ($\chi^2=4$) in this scenario by varying 
one parameter at a time  taking all the observables in {\tt Binned} case
 with unpolarized beams. The limits are presented  in the last 
column of Table~\ref{tab:Limits_Operator} for operator couplings $c_i^{\cal O}$  and  
 corresponding translated limit to the Lagrangian
couplings  $c_i^{{\cal L}_g}$ in Table~\ref{tab:Limits-gauged} using relation
from Eq.~\ref{eq:Operator-to-Lagrangian}.
Our one parameter limits are much better in comparison to the one parameter  tightest 
limits available from experiment, see Table~\ref{tab:aTGC_constrain_form_collider}.
 We calculate simultaneous limits using MCMC on the
operator couplings $c_i^{\cal O}$  and find  
the translated limits on the dependent Lagrangian couplings $c_i^{{\cal L}_g}$.
 The $95\%$ posterior simultaneous 
limits for five chosen beam polarization along $\eta_3=-\xi_3$, as in the previous subsection,  
are presented in Table~\ref{tab:Limits_Operator} on $c_i^{\cal O}$ and in 
Table~\ref{tab:Limits-gauged} on translated $c_i^{{\cal L}_g}$ for both {\tt Binned} and 
{\tt Unbinned} case. 
The pictorial representation of the limits 
on $c_i^{\cal O}$ and $c_i^{{\cal L}_g}$ are  presented in
%Tables~\ref{tab:Limits_Operator} and Table~\ref{tab:Limits-gauged} are shown in
Fig.~\ref{fig:Limits-Gauged} on the upper panels and 
lower panels, respectively. 

Here all the couplings do not vary in the same way over beam polarization (see Fig.~\ref{fig:Limits-Gauged}) in
contrast to the case of vertex factor approach in the previous subsection (see Fig.~\ref{fig:Limits}) .
 Based on the volume of limit the $(+0.8,-0.8)$ polarization provides best limits
 in the {\tt Unbinned} case while $(-0.8,+0.8)$ provides best limits in the {\tt Binned} 
case on couplings in this scenario on average. 
In comparison to the  one parameter tightest limit on couplings available in
literature from experiments (given in Table~\ref{tab:aTGC_constrain_form_collider}),
our simultaneous limits on the couplings given in Table~\ref{tab:Limits_Operator}, and in 
Table~\ref{tab:Limits-gauged} are comparable and better in some cases.
In the next section we  investigate on 
the choice of beam polarization based on minimum averaged likelihood.

%%%%%%%%%%%%%%%%%%%%%%%%%%%%%%%%%%%%%%%%%%%%%%%%%%%%%%%%%%%%%%%%%%%%%%%%%%%%%%%%%%%%%%%%%%%%%%%%%%%%%%%%
%%%%%%%%%%%%%%%%%%%%%%%%%%%%%%%%%%%%%%%%%%%%%%%%%%%%%%%%%%%%%%%%%%%%%%%%%%%%%%%%%%%%%%%%%%%%%%%%%%%%%%%%%
\section{On the choice of beam polarizations}\label{sec:4}

\begin{figure}[!t]
\centering
\includegraphics[width=8.50cm]{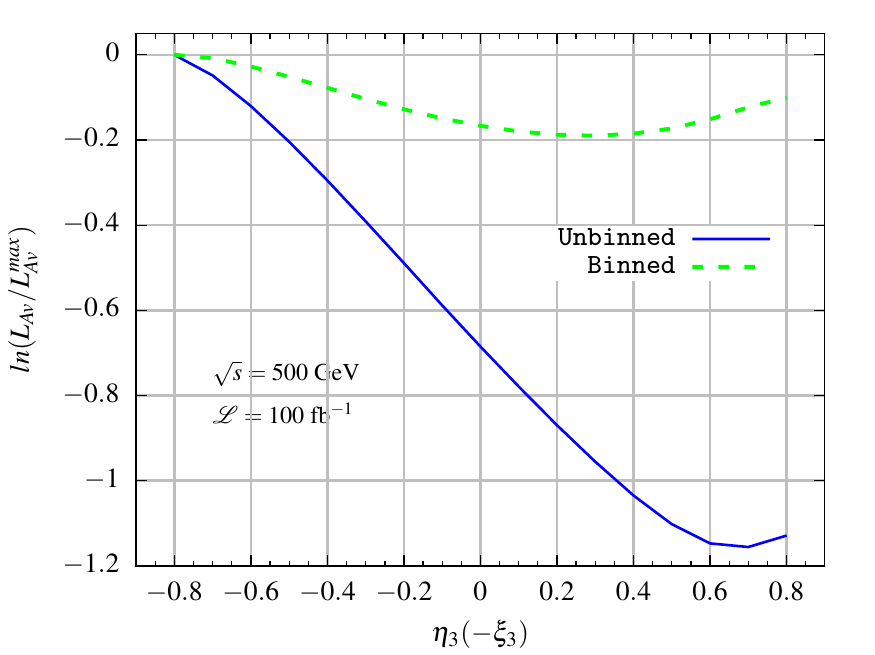}
\caption{\label{fig:MCInt} The normalized averaged Likelihood 
$L_{Av}=L(V_{\vec{f}},{\{\cal O}\};\eta_3,\xi_3)$ (normalized to $1$) 
in log scale as a function of $\eta_3 (-\xi_3)$ in $e^+e^-\to W^+W^-$ in the
Lagrangian approach for $\sqrt{s}=500$ GeV, ${\cal L}=100$ fb$^{-1}$. The {\it solid} (blue) line
and {\it dashed } (green) line represent the  {\tt Unbinned} and {\tt Binned} case, respectively} 
\end{figure}
\begin{figure}[!htb]
\centering
\includegraphics[width=8.5cm]{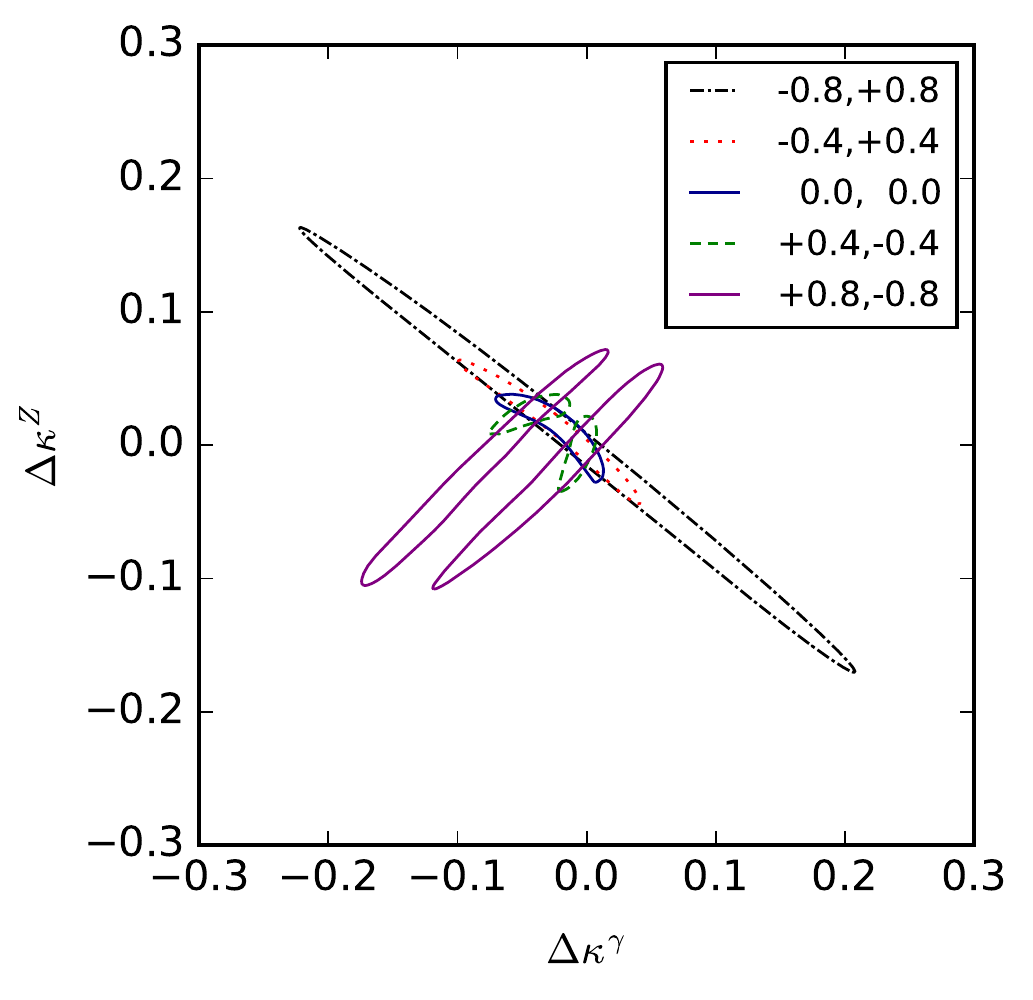}
\caption{\label{fig:getdistkakz} Two dimensional posterior $95\%$ BCI
contour of $\Delta\kappa^\gamma$ and $\Delta\kappa^Z$ in  $e^+e^-\to 
W^+W^-$ for $\sqrt{s}=500$ GeV, ${\cal L}=100$ fb$^{-1}$ for $5$ 
chosen set of longitudinal beam polarizations of $e^-$ and $e^+$  }
\end{figure}

In the previous section, we used the volume of the $95\%$ limits as a measure of
goodness of the combined limits to obtain the best choice of beam polarization.
Here we discuss the average likelihood or the
weighted volume of the parameter space define as~\cite{Rahaman:2017qql}
\begin{eqnarray}\label{eq:average_likelihood}
L(V_{\vec{f}},{\{\cal O}\};\eta_3,\xi_3)&=&\int_{V_{\vec{f}}}  
\exp{\bigg[-\frac{1}{2}\sum_{i} {\cal S}({\cal O}_i(\vec{f},\eta_3,\xi_3))^2 
\bigg]} d\vec{f} \nonumber\\
&=& \int_{V_{\vec{f}}} {\cal L}(\vec{f}, {\{\cal O}\};\eta_3,\xi_3) 
d\vec{f},
\end{eqnarray}
to cross-examine the beam polarization choices made in the previous section. 
Here $\vec{f}$ are the couplings and $V_{\vec{f}}$ is the volume 
of parameters space over which the average is done and has to be well above the volume of
limits. One naively expects the limits to be tightest when
$L(V_{\vec{f}},{\{\cal O}\};\eta_3,\xi_3)$ is minimum.
We calculate the above quantity as a function of $(\eta_3,\xi_3)$ 
along $\eta_3=-\xi_3$ for both the {\tt Unbinned} and {\tt Binned} case in the
effective vertex formalism given in Lagrangian in Eq.~\ref{eq:Lagrangian}. 
The normalized (normalized to $1$)  averaged 
likelihood as a function  $\eta_3 (-\xi_3)$ is shown in Fig.~\ref{fig:MCInt}.
We observe that the averaged likelihood curve does not follow the variation of limits 
over beam polarization presented in Table~\ref{tab:Limits} and Fig.~\ref{fig:Limits}, also
it does not have minima where limits are tightest. This is contrary to the naive
expectations.
This is because the region of the $14$ dimensional parameter space with
${\cal L}(\vec{f}, {\{\cal O}\};\eta_3,\xi_3)>e^{-25/2}$ (say the {\em blind
region}) is not a {\em convex hull}, i.e. the region of parameter space 
consistent with the SM at $5\sigma$ has a hole in it, like a $14$ dimensional 
hollow (or broken) ellipsoid, for $\eta_3>0$. As a result, the weighted volume 
of the blind region can become small while its size, the $1$-dimensional 
projections, remain large.
Even in the two dimensional projection of the blind region 
we see disconnected regions. This is shown in the two dimensional posterior
$95\%$ contour in $\Delta\kappa^\gamma - \Delta\kappa^Z$ plane in the {\tt 
Binned} case for five chosen polarizations in Fig.~\ref{fig:getdistkakz}.   
We see that an elliptical contour for beam polarization of $(-0.8,+0.8)$ 
(dotted/black) breaks into two disconnected regions for $(+0.4,-0.4)$ 
(solid/green) and then these regions grow in size for $(+0.8,-0.8)$ 
(dashed/purple).

\begin{figure}[!t]
\centering
\includegraphics[width=8.66cm]{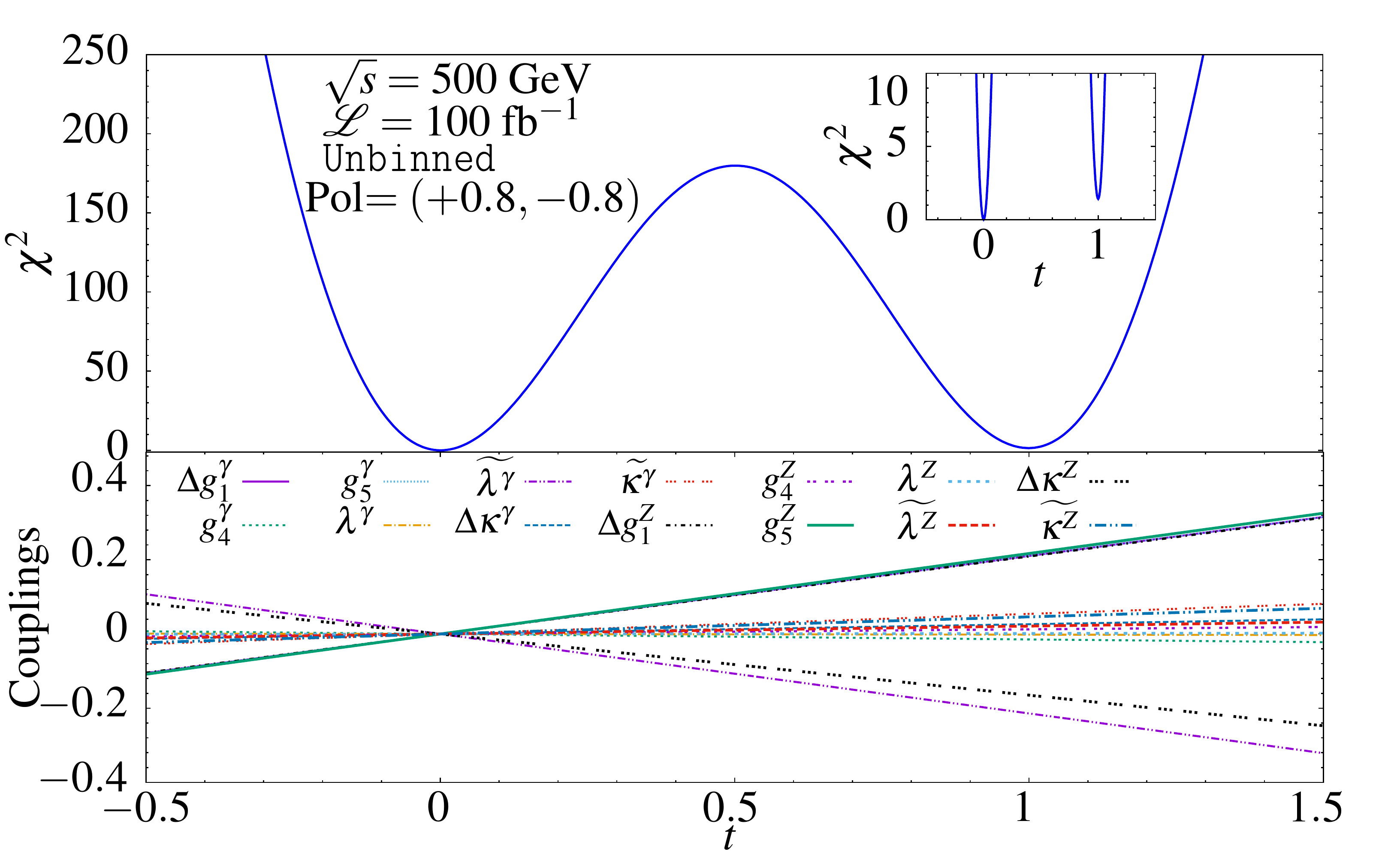}
\includegraphics[width=8.66cm]{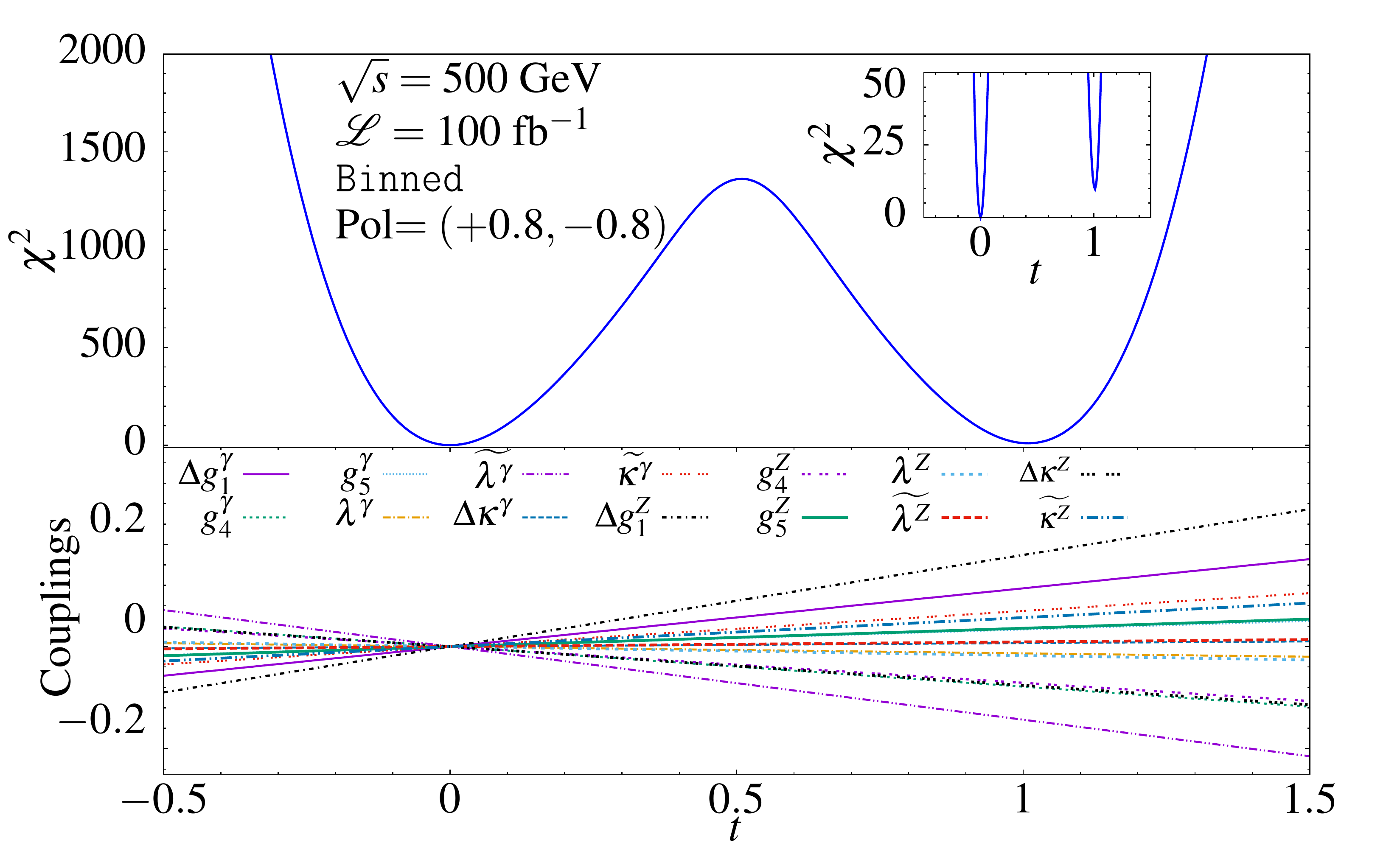}
\caption{\label{fig:chi2-double-hump} The  $\chi^2$ as a function of  
anomalous couplings $c_i^{\cal L}$  in the  chosen  direction shown in
the lower panel (see text for details)  for
{\tt Unbinned} case (above) and {\tt Binned} case (below) for the longitudinal
beam polarization $(+0.8,-0.8)$ of $e^-$ and $e^+$
in $e^+e^-\to W^+W^-$ for $\sqrt{s}=500$ GeV, ${\cal L}=100$ fb$^{-1}$. 
In the inset of two plots low $\chi^2$ are shown for visible distinguibility of two minima}
\end{figure}

To further illustrate the {\em non-convex} nature of the blind region we look
for regions where couplings $c_{i0}^{\cal L}$ are large but $\chi^2$ is low in
both {\tt Unbinned} and {\tt Binned} cases and show the variation of $\chi^2$ 
along the line joining the SM point and the point $c_{i0}^{\cal L}$. The
couplings along this line are parametrized as:
\begin{equation}
c_i^{\cal L}=t \times c_{i0}^{\cal L},~~i=1,2,...14,
\end{equation}
giving us the SM point for $t=0$ and the point $c_{i0}^{\cal L}$ for $t=1$.
The variation of $\chi^2$ as a function of $t$ is shown in
Fig.~\ref{fig:chi2-double-hump} for {\tt Unbinned} and {\tt Binned} cases
at $(+0.8,-0.8)$ beam polarization. 
We see that, for the {\tt Unbinned} case (Fig.~\ref{fig:chi2-double-hump}, top),
the regions with $t\in[0.11,0.88]$ has $\chi^2>25$ and hence outside the blind
region. That is the blind region for the {\tt Unbinned} case is not a convex
hull. Similarly, for the  {\tt Binned} case (Fig.~\ref{fig:chi2-double-hump},
bottom) the region with $t\in[0.04,0.97]$ are outside the blind region. This
non-convex shape of the blind region leads to a small value of
$L(V_{\vec{f}},{\{\cal O}\};\eta_3,\xi_3)$ while the size of the blind region
remains large.
%%%%%%%%%%%%%%%%%%%%%%%%%%%%%%%%%%%%%%%%%%%%%%%%%%%%%%%%%%%%%%%%%%%%%%%%%%%%%%%%%%%%%%%%%%
%%%%%%%%%%%%%%%%%%%%%%%%%%%%%%%%%%%%%%%%%%%%%%%%%%%%%%%%%%%%%%%%%%%%%%%%%%%%%%%%%%%%%%%%%%%

\section{Conclusion}\label{sec:conclusion}

In conclusion, we studied anomalous triple gauge boson couplings in 
$e^+e^-\to  W^+W^-$ with polarization observables of  $W$ boson
together with the total cross section and the forward backward asymmetry
for $\sqrt{s}=500$ GeV and integrated luminosity of ${\cal L}=100$ fb$^{-1}$
with polarization of $e^-$ and $e^+$ beams. We have $14$ anomalous couplings, where as we have 
only $10$ observables to measure them. So 
we  {\tt Binned} all the observables ($A_{fb}$ excluded) in $8$ regions of the 
$\cos\theta$ of $W$ to increase the number of observables to measure
the couplings. We estimated simultaneous limit on all the couplings 
in {\tt Unbinnded} case  as well as in {\tt Binned} case for several chosen set of 
beam polarization. We find $(+0.4,-0.4)$ to be the best beam polarization to 
obtain tightest constrain on the anomalous couplings $c_i^{\cal L}$  in both  {\tt Unbinned} 
and {\tt Binned} cases and {\tt Binned} limits are 
tighter (roughly twice) than the limits in  {\tt Unbinned} in each polarizations in the 
effective vertex factor formalism given in the Lagrangian in Eq.~\ref{eq:Lagrangian}. We further
estimated limits on anomalous couplings $c_i^{\cal O}$ from $SU(2)\times U(1)$
higher dimension effective  operators  in both  {\tt Unbinned} and 
{\tt Binned} cases for the same set of chosen beam polarizations and the 
translated limits on the non vanishing Lagrangian couplings $c_i^{{\cal L}_g}$. 
The limits on Lagrangian couplings in $SU(2)\times U(1)$ are tighter
than the limit when all $14$ couplings considered. We show the inconsistency between
best choice of beam polarizations and minimum likelihood averaged over the anomalous
couplings. This is because the blind region in non-convex resulting in small weighted volume, 
while the rectangular volume of 
parameters space from the limits is large. The best choice of beam polarization will be governed
solemnly based on the MCMC posterior limits on anomalous couplings, which is $(+0.4,-0.4)$
in the case of Lagrangian approach (can be seen in Table~\ref{tab:Limits} and Fig.~\ref{fig:Limits}).
For effective operator formalism the best choice is $(-0.8,+0.8)$ on the basis
of volume of rectangle formed by the limits can be seen in Tables~\ref{tab:Limits_Operator},
\ref{tab:Limits-gauged}. All the couplings do not poses tighter limits at this couplings,
some of them poses tighter limit on other choice of beam polarizations. Our one parameter
limits on $g_4^Z,~g_5^Z$ from Table~\ref{tab:Limits} and others from 
Tables~\ref{tab:Limits_Operator}, \ref{tab:Limits-gauged} are much better than the
available tightest one parameter limits from experiments in 
Table~\ref{tab:aTGC_constrain_form_collider}, while simultaneous limits are comparable. 

In the $W^\pm Z$ production at LHC, only the  $W^+W^-Z$ couplings appear limiting the 
number of anomalous couplings to $7$. With small number of couplings multivaluedness
may be avoided and hence tighter limit on anomalous couplings is expected at high
energy. The work on anomalous couplings in $W^\pm Z$ production at LHC is underway and
will be presented elsewhere.

%%%%%%%%%%%%%%%%%%%%%%%%%%%%%%%%%%%%%%%%%%%%%%%%%%%%%%%%%%%%%%%%%%%%%%%%%%%%%%%
%%%%%%%%%%%%%%%%%%%%%%%%%%%%%%%%%%%%%%%%%%%%%%%%%%%%%%%%%%%%%%%%%%%%%%%%%%%%%%%
%%%%%%%%%%%%%%%%%%%%%%%%%%%%%%%%%%%%%%%%
\noindent \textbf{Acknowledgements:} R.R. thanks Department of Science 
and Technology, Government of India for support through DST-INSPIRE Fellowship 
for doctoral program, INSPIRE CODE IF140075, 2014.  
%%%%%%%%%%%%%%%%%%%%%%%%%%%%%%%%%%%%%%%%%%%%%%%%%%%%%%%%%%%%%%%%%%%%%%%%%%%%%%%%%%%%%%%%%%%%%%%%%%%
%%%%%%%%%%%%%%%%%%%%%%%%%%%%%%%%%%%%%%%%%%%%%%%%%%%%%%%%%%%%%%%%%%%%%%%%%%%%%%%%%%%%%%%%%%%%%%%%%%%%%%
\appendix
%\begin{widetext}
\section{Labelling of anomalous  gauge boson couplings  and the dependences of observables  on
 anomalous gauge boson couplings $c_i^{\cal L}$ in $e^+e^-\to W^+W^-$ }\label{apendix:a}
The anomalous gauge boson couplings $c_i^{\cal O}$ of effective operator  
in Eq.~\ref{eq:opertaors5} and  the couplings $c_i^{\cal L}$ of the  
Lagrangian in Eq.~\ref{eq:Lagrangian}
and the couplings $c_i^{{\cal L}_g}$ of Lagrangian in $SU(2)\times U(1)$ (given in 
Eq.~\ref{eq:Operator-to-Lagrangian}) gauge are labelled as
\begin{eqnarray}
c_i^{\cal O}&=&\{ c_{WWW}, c_{W}, c_B, c_{\wtil{WWW}}, c_{\wtil{W}} \}\label{eq:ciO}\\
c_i^{\cal L}&=&\{ g_1^V,g_4^V,g_5^V,\lambda^V,\wtil{\lambda^V},\kappa^V, \wtil{\kappa^V} \},~~~ V=\gamma,Z \label{eq:ciL}\\
c_i^{{\cal L}_g}&=& \{ \lambda^V, \wtil{\lambda^V}, 
\Delta\kappa^\gamma, \wtil{\kappa^\gamma}, \Delta g_1^Z,  \Delta\kappa^Z, \wtil{\kappa^Z} \}\label{eq:ciLg}
\end{eqnarray}

\begin{table*}\caption{\label{tab:param_dependence} Dependences of observables 
 (numerators) on anomalous  couplings in the form of  $c_i^{\cal L}$ (linear), 
$(c_i^{\cal L})^2$ (quadratic) and $c_i^{\cal L}c_j^{\cal L},~i\ne j$ 
(interference)   in the process  $e^+e^-\to W^+W^-$. For the linear and 
quadratic terms $V=\gamma/Z$ and for cross interference terms $V=\gamma,Z$.
The ``\checkmark" ({\it checkmark}) represents the presence  and ``---" ({\it big-dash})
corresponds to  absence  of the term in the first column in the same row}
	\renewcommand{\arraystretch}{1.50}
	\begin{tabular*}{\textwidth}{@{\extracolsep{\fill}}llllllllllll@{}}\hline
		Parameters & $\sigma$ & $\sigma\times A_x$ & $\sigma\times A_y$ & $\sigma\times A_z$ &$\sigma\times A_{xy}$  
		&$\sigma\times A_{xz}$  & $\sigma\times A_{yz}$ & $\sigma\times A_{x^2-y^2}$ & $\sigma\times A_{zz}$ & $\sigma\times A_{fb}$ \\\hline
		$\Delta g_1^V$ & \checkmark & \checkmark & --- & \checkmark & --- & \checkmark & --- & \checkmark & \checkmark & \checkmark \\
		$g_4^V $& --- & --- & \checkmark & --- & \checkmark & --- & \checkmark & --- & --- & --- \\
		$g_5^V $& \checkmark & \checkmark & --- & \checkmark & --- & \checkmark & --- & \checkmark & \checkmark & \checkmark \\
		$\lambda^V $& \checkmark & \checkmark & --- & \checkmark & --- & \checkmark & --- & \checkmark & \checkmark & \checkmark \\
		$\wtil{\lambda^V}$ & --- & --- & \checkmark & --- & \checkmark & --- & \checkmark & --- & --- & --- \\
		$\Delta\kappa^V$ & \checkmark & \checkmark & --- & \checkmark & --- & \checkmark & --- & \checkmark & \checkmark & \checkmark \\
		$\wtil{\kappa^V}$ & --- & --- & \checkmark & --- & \checkmark & --- & \checkmark & --- & --- & --- \\
		$(\Delta g_1^V)^2 $& \checkmark & \checkmark & --- & --- & --- & --- & --- & \checkmark & \checkmark & --- \\
		$(g_4^V)^2$& \checkmark & --- & --- & --- & --- & --- & --- & \checkmark & \checkmark & --- \\
		$(g_5^V)^2 $& \checkmark & --- & --- & --- & --- & --- & --- & \checkmark & \checkmark & --- \\
		$(\lambda^V)^2$& \checkmark & \checkmark & --- & --- & --- & --- & --- & \checkmark & \checkmark & --- \\
		$(\wtil{\lambda^V})^2$& \checkmark & \checkmark & --- & --- & --- & --- & --- & \checkmark & \checkmark & --- \\
		$(\Delta\kappa^V)^2 $& \checkmark & \checkmark & --- & --- & --- & --- & --- & \checkmark & \checkmark & --- \\
		$(\wtil{\kappa^V})^2 $& \checkmark & \checkmark & --- & --- & --- & --- & --- & \checkmark & \checkmark & --- \\
		$\Delta g_1^V g_4^V $& --- & --- & --- & --- & --- & --- & \checkmark & --- & --- & --- \\
		$\Delta g_1^V g_5^V $& --- & --- & --- & \checkmark & --- & --- & --- & --- & --- & \checkmark \\
		$\Delta g_1^V \lambda^V $& \checkmark & \checkmark & --- & --- & --- & --- & --- & \checkmark & \checkmark & --- \\
		$\Delta g_1^V \wtil{\lambda^V} $& --- & --- & \checkmark & --- & \checkmark & --- & --- & --- & --- & --- \\
		$\Delta g_1^V \Delta\kappa^V $& \checkmark & \checkmark & --- & --- & --- & --- & --- & \checkmark & \checkmark & --- \\
		$\Delta g_1^V \wtil{\kappa^V} $& --- & --- & \checkmark & --- & \checkmark & --- & --- & --- & --- & --- \\
		$g_4^V g_5^V $& --- & --- & --- & --- & \checkmark & --- & --- & --- & --- & --- \\
		$g_4^V \lambda^V $& --- & --- & --- & --- & --- & --- & \checkmark & --- & --- & --- \\
		$g_4^V \wtil{\lambda^V} $& --- & --- & --- & \checkmark & --- & \checkmark & --- & --- & --- & \checkmark \\
		$g_4^V \Delta\kappa^V $& --- & --- & --- & --- & --- & --- & \checkmark & --- & --- & --- \\
		$g_4^V \wtil{\kappa^V} $& --- & --- & --- & \checkmark & --- & \checkmark & --- & --- & --- & \checkmark \\
		$g_5^V \lambda^V $& --- & --- & --- & \checkmark & --- & \checkmark & --- & --- & --- & \checkmark \\
		$g_5^V \wtil{\lambda^V} $& --- & --- & --- & --- & --- & --- & \checkmark & --- & --- & --- \\
		$g_5^V \Delta\kappa^V $& --- & --- & --- & \checkmark & --- & \checkmark & --- & --- & --- & \checkmark \\
		$g_5^V \wtil{\kappa^V} $& --- & --- & --- & --- & --- & --- & \checkmark & --- & --- & --- \\
		$\lambda^V \wtil{\lambda^V}$ & --- & --- & \checkmark & --- & \checkmark & --- & --- & --- & --- & --- \\
		$\lambda^V \Delta\kappa^V $& \checkmark & \checkmark & --- & --- & --- & --- & --- & \checkmark & \checkmark & --- \\
		$\lambda^V\wtil{\kappa^V}  $& --- & --- & \checkmark & --- & \checkmark & --- & --- & --- & --- & --- \\
		$\wtil{\lambda^V} \Delta\kappa^V $& --- & --- & \checkmark & --- & \checkmark & --- & --- & --- & --- & --- \\
		$\wtil{\lambda^V}\wtil{\kappa^V}  $& \checkmark & \checkmark & --- & --- & --- & --- & --- & \checkmark & \checkmark & --- \\
		$\Delta\kappa^V \wtil{\kappa^V} $& --- & --- & \checkmark & --- & \checkmark & --- & --- & --- & --- & --- \\
		\hline
	\end{tabular*}
\end{table*}

%%%%%%%%%%%%%%%%%%%%%%%%%%%%%%%%%%%%%%%%%%%%%%%%%%%%%%%%%%%%
%\newpage
%\fontsize{11}{9}\selectfont
\fontsize{9.5}{11}\selectfont
\bibliography{Bibliography}
\bibliographystyle{utphys}
\end{document}